\documentclass[reprint,aps,pre]{revtex4-2}

\usepackage[T1]{fontenc} 
\usepackage[utf8]{inputenc} 
\usepackage[english]{babel}		
\usepackage{amsmath}	
\usepackage{amssymb}
\usepackage{graphicx} 
\usepackage{hyperref}
\usepackage[]{cleveref}

\Crefname{equation}{Eq.}{Eqs.}

\newcommand{\be}{\begin{equation}}
\newcommand{\ee}{\end{equation}}

\newcommand{\nn}{\nonumber}
\newcommand{\lp}{\left(}
\newcommand{\rp}{\right)}
\newcommand{\lbk}{\left[}
\newcommand{\rbk}{\right]}
\newcommand{\di}{\mathrm{d}}
\newcommand{\T}[1]{\text{#1}}
\newcommand{\Var}{\mathrm{Var}}

\begin{document}
	
\title{Universal constraints on selection strength in lineage trees}
\author{Arthur Genthon}
\author{David Lacoste}
\affiliation{Gulliver UMR CNRS 7083, ESPCI Paris, Université PSL, 75005 Paris, France}
\email{arthur.genthon@espci.fr}

\begin{abstract} 
We obtain general inequalities constraining the difference between the average of an arbitrary function of a phenotypic trait, which includes the fitness landscape of the trait itself, in the presence or in the absence of natural selection. These inequalities imply bounds on the strength of selection, which can be measured from the statistics of trait values and divisions along lineages. The upper bound is related to recent generalizations of linear response relations in Stochastic Thermodynamics, and shares common features with Fisher's fundamental theorem of natural selection, and with its generalization by Price, although they define different measures of selection. The lower bound follows from recent improvements on Jensen's inequality, and both bounds depend on the variability of the fitness landscape. We illustrate our results using numerical simulations of growing cell colonies and with experimental data of time-lapse microscopy experiments of bacteria cell colonies.
\end{abstract}

\maketitle

\section{Introduction}
\label{sec_intro}

Quantifying the strength of selection in populations is an essential step in any description of evolution. 
With the development of single cell measurements, a large amount of data on cell lineages is becoming available both at the genotypic and phenotypic level. By analyzing the statistics of cell divisions in population trees, one can measure selection more accurately than using classical population growth rate measurements \cite{leibler_individual_2010}. Similarly, by tracking phenotypes on cell lineages, one can obtain statistically reliable estimations of the fitness landscape of a given trait and of the selection strength of that trait  \cite{nozoe_inferring_2017}. In addition, an optimal lineage principle can be used to infer the population growth rate \cite{wakamoto_optimal_2012} or selective forces \cite{lambert_quantifying_2015} from lineage statistics. All these methods contribute to bridging the gap between single-cell experiments at the population level and molecular mechanisms \cite{lin_effects_2017}.

An alternate method to infer selection in evolution focuses on dynamical trajectories of frequency distributions  \cite{mustonen_fitness_2010,mustonen_fitness_2009}. In these works, Mustonen et al. introduced the notion of fitness flux to characterize the adaptation of a population by taking inspiration from Stochastic Thermodynamics. In fact, ideas from Stochastic Thermodynamics can be applied directly at the level of individual cell trajectories \cite{kobayashi_fluctuation_2015}. By following this kind of approach, we have derived general constraints on dynamical quantities characterizing the cell cycle such as the average number of divisions or the mean generation time \cite{garcia-garcia_linking_2019,genthon_fluctuation_2020}. These constraints are universal because they hold independently of the specific cell dynamic model and they are indeed verified in experimental data.
Other examples of universality in the context of evolution includes the identification of universal families of distributions of selected values and the use of methods from extreme value statistics \cite{boyer_hierarchy_2016,smerlak_limiting_2017}.
 
Here, we derive universal constraints for the average value of a function of a trait, and for its selection strength, by exploiting a set of recent results known under the name of Thermodynamic Uncertainty Relations (TUR).
These relations take the form of inequalities, which generalize fluctuation-response relations far from equilibrium \cite{dechant_fluctuationresponse_2020}, and which capture important trade-offs for thermodynamic and non-thermodynamic systems \cite{dinis_phase_2020} as recently reviewed in \cite{horowitz_thermodynamic_2020}. 
Although our results are framed in the context of cell population in lineage trees, they apply more broadly to general stochastic processes defined on any branched tree.

We start in \cref{sec_framework} by laying the theoretical framework and the definitions of the forward and backward samplings of lineages within a tree, which are at the core of the notions of fitness landscape and selection strength. In \cref{sec_ineq}, we derive a general upper bound for the difference between the average values of an observable with respect to two different probability distributions, which we use in \cref{sec_ineq_pi} to obtain an upper bound for the strength of selection. That result goes beyond the Gaussian approximation. In \cref{sec_small_var}, we study the case of small variability which leads to a simple expression of the strength of selection, reminiscent of the Gaussian case. Those expressions have mathematical similarities with Fisher's fundamental theorem of natural selection and Price's equation, although they correspond to different definitions of selection, as detailed in \cref{sec_fisher_price}. To complement the upper bound on the strength of selection, we use a recent sharpened version of Jensen's inequality to derive in \cref{sec_lower_bbound} a lower bound for the strength of selection. Both bounds are tested with simulations and experimental data in \cref{sec_test}, showing a very good agreement with the theory.
Finally, we conclude in \cref{sec_discussion}. Several appendices (\crefrange{sec_app_h_vs_lambda}{sec_app_h_age}) present the details of the calculations, supplementary figures, and numerical comparisons between our results and results previously published.

\section{A framework for lineage statistics}
\label{sec_framework}

A colony of cells can be represented as a branched tree, whose branches are called lineages and whose nodes correspond to cell divisions. We assume that each cell in the population divides after a stochastic time into exactly $m$ daughter cells. In order to extract relevant statistics from such a tree, one needs to sample the lineages following a weighting scheme. 
The backward (or retrospective) and the forward (or chronological) samplings have been introduced in the context of populations of cells \cite{nozoe_inferring_2017,thomas_making_2017,lin_effects_2017}, and previously defined in the mathematical literature \cite{baake_mutation_2007,smerlak_quasi-species_2021}. 
The backward sampling of lineages assigns a uniform weight $N(t)^{-1}$ to each of the $N(t)$ lineages, leading to an over-representation of cells coming from sub-populations that divided more than average. To compensate this bias, the forward sampling takes into account the number of divisions $K$ along a lineage and assigns to the lineage a weight $N_0^{-1} m^{-K}$, where $N_0$ is the size of the initial population. 
Intuitively, a lineage is followed forward in time from a cell in the initial population, by choosing with uniform probability $1/m$ which daughter to follow among the $m$ daughters at each division. In this sense, the forward sampling cancels the effect of selection because the sister cells born from the same division have the same weight, regardless of their reproductive successes, i.e. the sizes of the sub-populations they generate. Thus, the statistics obtained with a forward sampling of the lineages within a tree reproduces the statistics obtained in single-lineage experiments, like in mother-machine configuration \cite{wang_robust_2010}.
A graphical example of the two samplings for a simple tree is given on \cref{fig_price_vs_us}.

A general phenotypic trait $\mathcal{S}$ then admits a forward and a backward distributions respectively defined by $p_{\T{for}} (s,t)= \sum_{K=0}^{\infty} n(s,K,t)/(N_0 m^K)$ and $p_{\T{back}} (s,t)= n(s,t)/N(t)$, where $n(x,t)$ is the number of lineages featuring a cell with trait value $x$ at time $t$.
Comparing $p_{\T{for}} (s,t)$ and $p_{\T{back}} (s,t)$ offers some insight on the effect of selection on trait $\mathcal{S}$.
For this purpose, we define the fitness landscape as \cite{nozoe_inferring_2017}
\begin{equation}
\label{eq_def_fit_land_s}
h_t(s)=\Lambda_t + \frac{1}{t} \ln \lbk \frac{p_{\T{back}} (s,t)}{p_{\T{for}} (s,t)} \rbk \,,
\end{equation} 
where $\Lambda_t= \ln (N(t) /N_0) /t $ is the population growth rate. 
Note that in the classical framework of evolutionary dynamics, the notion of fitness landscape finds its origin in Wright's seminal work \cite{wright_roles_1932}, and is defined as a mapping between the values or versions of a phenotype or a genotype, with their associated fitnesses \cite{peliti_introduction_1997}. The notion of fitness in biology has multiple meanings, but is often understood in this context as the reproductive success, or growth rate. In contrast, this is not the case for the fitness landscape we defined, which therefore should not be confused with the growth rate of the sub-population carrying the trait value $s$.
Indeed, the reproductive success is defined by the comparison of the frequencies of a trait in a population over time, whereas the fitness landscape as we defined it compares the frequencies of a trait at the same time but in ensembles with and without selection. Therefore, $h_t(s_1)>h_t(s_2)$ means that the trait value $s_1$ benefits more from selection than $s_2$, but not necessarily that $s_1$ has a greater reproductive success than $s_2$. As a consequence, cells carrying the value $s_1$ could still be less represented in the population than those carrying trait $s_2$. The two points of view are linked by simple relations as detailed in \cref{sec_app_h_vs_lambda}, and we argue in \cref{sec_fisher_price} that this subtle difference leads to different definitions of the effect of selection, and that the point of view which compares chronological and retrospective distributions could be more suitable to describe selection for certain applications.

When the statistics of trait $\mathcal{S}$ is unaffected by selection, that is when there is no correlation between the number of divisions undergone by a cell and the value $s$ for this trait, then $p_{\T{back}} (s,t) = p_{\T{for}} (s,t)$ and the fitness landscape is flat, equal to the population growth rate. 
Instead, if the statistics of the trait is strongly perturbed by selection, then the fitness landscape is more rough and exhibits important deviations from its mean value. 

Therefore, the variance of the fitness landscape appears as a natural candidate to quantify the roughness of this fitness landscape effect. However, this variance can be computed in both the forward or backward ensembles, giving related but different results, and it is therefore unclear which of the two should be used. 
To resolve this issue, we define the strength of selection $\Pi_{\mathcal{S}}$ acting on the trait $\mathcal{S}$ as the change in mean fitness landscape between the ensembles with and without selection \cite{nozoe_inferring_2017}:
\begin{equation}
\label{eq_def_pi}
\Pi_{\mathcal{S}}=\langle h_t(s) \rangle_{\T{back}} - \langle h_t(s) \rangle_{\T{for}} \,.
\end{equation}
This quantity indeed reflects the roughness of the fitness landscape, since it is null when the fitness landscape is flat and becomes larger as the difference between the backward and forward statistics for trait $\mathcal{S}$ increases. This behavior is well-understood by writing the strength of selection as \cite{nozoe_inferring_2017}:
\begin{equation}
\label{eq_def_pi_jef}
\Pi_{\mathcal{S}}=\frac{1}{t} \mathcal{J}(p_{\T{back}} (s,t) | p_{\T{for}} (s,t)) \,,
\end{equation}
where $\mathcal{J}$ is the Jeffrey's divergence, a non-negative and symmetric information-theoretic distance between the two distributions $p_{\T{back}} (s,t)$ and $p_{\T{for}} (s,t)$, defined as $\mathcal{J}(p(x) | q(x)) = \int (p(x)-q(x)) \ln(p(x)/q(x)) \di x$.

Let us briefly comment on two points.
First, the strength of selection defined here should not be confused with the coefficient of selection, usually defined as the relative difference in fitness associated with two values of a phenotypic trait \cite{mustonen_fitness_2010}. 
Second, the strength of selection is a function of time, since fitness landscapes are time-dependent by definition. 
Only if a steady state is reached in the long time limit, then $h_t(s)$ tends to a constant equal to the steady state population growth rate $\Lambda$, and the strength of selection tends to $0$, as expected since selection no longer shifts trait frequencies.

In the particular case of Gaussian distributions, the strength of selection $\Pi_{\mathcal{S}}$ and the variance of the fitness landscape are in fact linked by a very simple relation. More precisely, when the forward distribution for the fitness landscape is Gaussian, and for a bijective function $h_t(s)$, then its backward distribution is also Gaussian, with the same standard deviation but a shifted average value, leading to (see \cref{sec_app_gauss})
\begin{equation}
\label{eq_hs_back_gauss}
\Pi_{\mathcal{S}}=t \ {\rm Var}(h_t) \,,
\end{equation}
where the variance can be indifferently taken over the forward or backward sampling. 
Note that we recover here a result known from \cite{nozoe_inferring_2017}, in a more direct way and with restricted assumptions, since in that reference the authors derived this relation assuming that the joint distribution of $h_t(s)$ and $K$ was a bivariate Gaussian distribution.

However, the Gaussian case only covers a small portion of realistic cases, and fitness landscapes can exhibit strong deviations from Gaussian distributions. In the context of age-controlled divisions \cite{genthon_fluctuation_2020}, it can be shown that the distribution of age fitness landscape is non Gaussian and depends on the shape of the division rate as a function of the age. Moreover, we show on \cref{fig_hx} in \cref{sec_test} (and on \cref{fig_app_ha} in \cref{sec_app_h_age}) experimental fitness landscapes that are non Gaussian. 

In this article, we derive universal relations going beyond the Gaussian assumption, and obtain a set of upper and lower bounds for the strength of selection, in terms of both the forward and backward variances for the fitness landscape. 
To do so, let us first derive a general inequality constraining the difference in average value for an observable between two probability distributions.

\section{General fluctuation-response inequality}
\label{sec_ineq}

We consider a general system described by a reference probability distribution $p_{\T{a}} (s,t)$, where $s$ is the value taken by a state variable $\mathcal{S}$. By perturbing the system, we change the distribution of the variable $\mathcal{S}$ from $p_{\T{a}} (s,t)$ to $p_{\T{b}} (s,t)$. We consider an observable depending on the variable $\mathcal{S}$, through a function $g_t(s)$, and ask the question of how the mean value of this observable is modified when the system is perturbed. 

Assuming that $p_{\T{a}} (s,t)$ and $p_{\T{b}} (s,t)$ have the same support, we can define the ratio 
\begin{equation}
\label{eq_def_q}
q_t(s)=\frac{p_{\T{b}} (s,t)}{p_{\T{a}} (s,t)} \,.
\end{equation}
Let us now compute the covariance between $g_t(s)$ and $q_t(s)$ with respect to $p_{\T{a}} (s,t)$:
\begin{align}
\label{eq_cov_gen}
{\rm{Cov}}_{\rm{a}} (g_t,q_t) &= \langle g_t q_t \rangle_{\T{a}}  - \langle g_t \rangle_{\T{a}} \langle q_t \rangle_{\T{a}} \nn \\
&=\langle g_t \rangle_{\T{b}}-\langle g_t \rangle_{\T{a}} \,,
\end{align}
where we used $\langle q_t \rangle_{\T{a}}=1$, due to the normalization of $p_{\T{b}}$, and  $\langle q_t g \rangle_{\T{a}}=\langle g_t \rangle_{\T{b}}$.

Following the method used in \cite{dinis_phase_2020} to derive mean-variance trade-off bounds in horse race gambling, we use the Cauchy-Schwarz inequality for the covariance:
\begin{equation}
\label{eq_CS_cov}
{\rm{Cov}}_{\rm{a}} (g_t,q_t)^2 \leq \sigma^2_{\T{a}}(g_t) \sigma^2_{\T{a}}(q_t) \,,
\end{equation}
with $\sigma^2_{\T{a}}$ the variance with respect to $p_{\T{a}} (s,t)$. 
Finally, by combining \cref{eq_cov_gen,eq_CS_cov}, we obtain a general bound for the difference in average values:
\begin{equation}
\label{eq_tur_s}
| \langle g_t \rangle_{\T{b}}-\langle g_t \rangle_{\T{a}}| \leq \sigma_{\T{a}}(g_t) \sigma_{\T{a}}(q_t) \,.
\end{equation}
%

The inequality can be understood as an out-of-equilibrium generalization of the fluctuation-dissipation theorem, because it involves a comparison between a reference unperturbed dynamics and a perturbed dynamics. The difference between the unperturbed and the perturbed averages of the function $g_t(s)$ is bounded by the unperturbed fluctuations of this function, measured by $\sigma_{\T{a}}(g_t)$, times $\sigma_{\T{a}}(q_t)$ which is a information-theoretic distance between the two probability distributions.
Indeed, since $\langle q_t \rangle_{\T{a}}=1$, the variance of $q_t$ is given by $\sigma^2_{\T{a}}(q_t)=\int_s \di s \ (p_{\T{b}}(s)-p_{\T{a}}(s))^2/p_{\T{a}}(s)$, and thus the larger $\sigma_{\T{a}}(q_t)$, the further away $p_{\T{b}} (s,t)$ and $p_{\T{a}} (s,t)$ are from each other. 

To derive \cref{eq_tur_s}, we adopted the point of view of the unperturbed statistics $p_{\T{a}} (s,t)$ as reference, but a similar bound can be obtained in terms of standard deviations with respect to the perturbed dynamics $p_{\T{b}} (s,t)$. 
We consider the covariance between  $g_t(s)$ and $r_t(s)=1/q_t(s)$, with respect to $p_{\T{b}} (s,t)$:
\begin{equation}
\label{eq_cov_gen_b}
{\rm{Cov}}_{\rm{b}} (g_t,r_t) =\langle g_t \rangle_{\T{b}}-\langle g_t \rangle_{\T{a}} \,,
\end{equation}
Following the same steps, and using the Cauchy-Schwarz inequality for this covariance we finally obtain
\begin{equation}
\label{eq_tur_s_back}
| \langle g_t \rangle_{\T{b}}-\langle g_t \rangle_{\T{a}}| \leq \sigma_{\T{b}}(g_t) \sigma_{\T{b}}(r_t) \,,
\end{equation}
where the term $\sigma_{\T{b}}(r_t)$ is similarly interpreted as an information-theoretic distance measure between the two distributions.
Thus, combining \cref{eq_tur_s,eq_tur_s_back}, the change in mean value of the function $g_t$ of the variable  $\mathcal{S}$ between an unperturbed and a perturbed statistics is bounded by
\begin{equation}
\label{eq_tur_s_min}
| \langle g_t \rangle_{\T{b}}-\langle g_t \rangle_{\T{a}}| \leq \min  \lp \sigma_{\T{a}}(g_t) \sigma_{\T{a}}(q_t), \sigma_{\T{b}}(g_t) \sigma_{\T{b}}(r_t) \rp \,.
\end{equation}

A similar bound for $| \langle g_t \rangle_{\T{b}}-\langle g_t \rangle_{\T{a}}| $  was derived by Dechant et al. in \cite{dechant_fluctuationresponse_2020}, using Jensen's inequality. Their bound (eq. 5 or 11 in their text) also involves a measure of the distance between the two probability distributions (Kullback-Leibler divergence) and the standard deviation of the observable considered in the unperturbed dynamics. 
We carry out a numerical comparison between the two bounds in \cref{sec_app_bounds_vs_sasa}, to find which one is the tightest of the two. This shows that the relative performance of the two bounds depends on the shape of the perturbed and unperturbed distributions. 
In any case, our bound is easy to evaluate since it does not require an optimization over a free parameter, as it is the case in \cite{dechant_fluctuationresponse_2020} (see \cref{eq_app_def_uds}).

\section{The strength of selection is bounded by the variability in fitness landscape}
\label{sec_ineq_pi}

The results derived in the previous section for general distributions $a$ and $b$ are now used to obtain constraints on the strength of selection. Indeed, by setting the unperturbed distribution $a$ to be the forward distribution of a phenotypic trait $\mathcal{S}$ and the perturbed distribution $b$ to be the backward distribution of this trait (which is allowed since the forward and backward distributions have the same support), the difference $ \langle g_t \rangle_{\T{back}}-\langle g_t \rangle_{\T{for}} $ is the change of mean value for $g_t$ between an ensemble without selection (forward) and with selection (backward), while the perturbation is measuring the selection itself.
In this context, the ratio $q_t(s)$ and the fitness landscape $h_t(s)$ are linked by the simple relation $q_t(s)=\exp \lbk t \lp h_t(s) - \Lambda_t \rp \rbk$.

An important application of the above results is when the arbitrary function $g_t(s)$ is the fitness landscape $h_t(s)$ itself. 
In this case, \cref{eq_cov_gen,eq_cov_gen_b} read
\begin{align}
\label{eq_pi_equality_cov}
\Pi_{\mathcal{S}} &= {\rm{Cov}}_{\rm{for}} (h_t, e^{t h_t}) \, e^{-t \Lambda_t}\\
&= {\rm{Cov}}_{\rm{back}} (h_t, e^{-t h_t}) \,  e^{t \Lambda_t} \,.
\end{align}
These equalities generalize the linear relation between the strength of selection and the variance of the fitness landscape, valid in the Gaussian case (\cref{eq_hs_back_gauss}). 
To better highlight the role of the variability of the fitness landscape, we write \cref{eq_tur_s_min} in this context
\begin{equation}
\label{eq_tur_hs}
\Pi_{\mathcal{S}}
\leq \min  \lp \sigma_{\T{for}}(h_t) \sigma_{\T{for}}(q_t), \sigma_{\T{back}}(h_t) \sigma_{\T{back}}(r_t) \rp \,, 
\end{equation}
where the absolute values in the l.h.s. can be removed because the strength of selection is defined positive, as deduced from \cref{eq_def_pi_jef}.

Note that that the l.h.s. of \cref{eq_tur_hs} involves averages with respect to the two probability distributions, unlike what happens in the standard TUR where only one such average is present. The reason is that in Stochastic Thermodynamics, the two relevant probability distributions correspond to a forward and a time-reversed dynamics, and the quantity which replaces $g_t(s)$ is a current, which changes sign under time reversal symmetry. Here there is no such symmetry present, hence the two averages are not the opposite of one another.

We obtained a universal upper bound for the strength of selection acting on trait $\mathcal{S}$, which involves the information-theoretic distances $\sigma_{\T{for}}(q_t)$ and $\sigma_{\T{back}}(r_t)$ between the backward and forward statistics, and the variances of the fitness landscape in both ensembles, which are in general different from each other.

Even if the interpretation of $\sigma_{\T{for}}(q_t)$ as a distance in the framework of linear-response theory is general, $\sigma_{\T{for}}(q_t)$ can also be expressed in terms of measurable quantities for cell colonies:
\begin{equation}
\label{eq_std_q}
\sigma_{\T{for}}(q_t)=\frac{\sigma_{\T{for}}(e^{t h_t})}{\langle e^{t h_t} \rangle_{\T{for}}} \,.
\end{equation}
Thus, $\sigma_{\T{for}}(q_t)$ quantifies the relative fluctuation of the quantity $\exp \lbk th_t(s) \rbk$, which itself represents the ratio of the expected number of lineages ending with trait value $s$, rescaled by the number $N_0$ of initial cells, to the forward probability of this trait value (see \cref{sec_app_std_q}). A similar interpretation can be given for the term $\sigma_{\T{back}}(r_t)$.

\section{A linear response equality}
\label{sec_small_var}

Let us now investigate precisely the conditions for which the previous inequalities become saturated.
It is straight-forward to show that when the forward and backward statistics are equal, inequalities \cref{eq_tur_hs,eq_tur_s_min} are saturated. Indeed, the l.h.s terms are $0$ and the r.h.s terms are null because they contain the standard deviation of the constant quantities $q_t(s)=r_t(s)=1$.

We now study the case where the two probability distributions approach each other. One possible measure of the distance between the two distributions is $\sigma(q)$, or equivalently $\sigma(\ln q)=t \sigma(h_t)$. In the limit $t \sigma(h_t) \to 0$, referred to as the small variability limit, the l.h.s of \cref{eq_tur_s_min} reads (see \cref{sec_app_small_var})
\begin{equation}
\label{eq_g_small_var}
\langle g_t \rangle_{\T{back}}-\langle g_t \rangle_{\T{for}} \underset{t \sigma \rightarrow 0}{\sim} t \ {\rm Cov}(h_t,g_t) \,,
\end{equation}
and the l.h.s of \cref{eq_tur_hs} when the function $g_t$ is the fitness landscape itself reads
\begin{equation}
\label{eq_pi_small_var}
\Pi_{\mathcal{S}} \underset{t \sigma \rightarrow 0}{\sim} t \ {\rm Var}(h_t) \,,
\end{equation}
where the variance and the covariance can be equivalently taken over the forward or backward sampling. 
When computing the r.h.s of \cref{eq_tur_hs,eq_tur_s_min}, we obtain that \cref{eq_tur_hs} is saturated in this limit whereas \cref{eq_tur_s_min} is not.
The limit can also be written $t \ll \sigma(h_t)^{-1}$ which defines a characteristic timescale of the system. 
In practice, this limit can be reached either for short times or in the case of a strong control mechanism on the divisions, leading the lineages to stay synchronized even after a finite time. It is also possible to regard this limit as a regime of weak selection \cite{neher_statistical_2011}, since the strength of selection is small precisely because of \cref{eq_tur_hs}.

\section{Comparison with Fisher's fundamental theorem and Price's equation}
\label{sec_fisher_price}

In this section, we highlight the similarities between our results and the relations derived by Fisher and Price, in which the population growth rate, or fitness, associated with a trait value $s$ plays a similar role to our fitness landscape $h_t(s)$. However, because these notions of fitness are distinct, as explained in \cref{sec_framework} and further analyzed in \cref{sec_app_h_vs_lambda}, the interpretations of selection contained in these equations are qualitatively different. 

Fisher's fundamental theorem of natural selection states that the time derivative of the mean fitness of a population is equal to the variance of the fitness across the population \cite{fisher_genetical_2000,price_fishers_1972}: $\di \Lambda_t / \di t = \Var_{\rm{back}}(\Lambda_i(s,t))$,
%
%
where $\Lambda_i(s,t)= (\di n(s,t)/\di t)/n(s,t)$ is the instantaneous growth rate, or instantaneous fitness, of the sub-population of size $n(s,t)$ carrying the trait value $s$. The variance is computed with respect to the backward distribution, which puts equal weights on individuals, and therefore is the natural distribution to consider.
The r.h.s of both \cref{eq_pi_small_var,eq_hs_back_gauss} and Fisher's theorem involve the variance of a certain kind of fitness within the population. In contrast, the l.h.s. in Fisher's theorem is a measure of evolution of the population, while the l.h.s. in our result is a trait-dependent measure of selection. 
Moreover, some well-known limitations of Fisher's theorem lie in the implicit assumption that natural selection is the only possible phenomenon leading to a change in the gene frequencies \cite{price_fishers_1972}. This assumption neglects many important phenomena such as mutations and recombination events \cite{neher_statistical_2011}, random drift due to finite population size, and specific features of seascapes \cite{mustonen_fitness_2009}. In contrast, our result does not suffer from any of these limitations, since it only requires the population to be represented as a branched tree, and is completely independent of the dynamics that generates the tree.

Price's equation \cite{price_fishers_1972} predicts the time evolution of the mean value of a trait, and involves two terms: a covariance term representing the selection effect, and the `environment change term', or dynamic effect, which accounts for all the other sources of variability leading to a change in the mean value of the trait. 
%
The part of the time evolution of $\langle s \rangle_{\T{back}}$ that is due to the separate effect of natural selection, in Price's sense which we denote with the superscript $\rm{NS}$, can then be written as 
\begin{equation}
\Delta \langle s \rangle_{\T{back}}^{\rm{NS}}= \frac{1}{\Lambda} \ \rm{Cov}_{\T{back}} (s,\Lambda(s)) \,,
\end{equation}
where $\Delta \langle s \rangle_{\T{back}}^{\rm{NS}}=(\langle s \rangle_{\T{back}}(t+\tau)-\langle s \rangle_{\T{back}}(t))^{\rm{NS}}$, $\Lambda(s)=n(s,t+\tau)/n(s,t)$, and $\Lambda=\langle \Lambda(s) \rangle_{\T{back}}$.
We can draw a parallel between this equation and \cref{eq_g_small_var}, as their r.h.s. both involve the covariance of the trait subjected to selection and a fitness associated to it.
Note that there is no `environment change term' in \cref{eq_g_small_var} because the strength of selection is defined precisely in such a way as to isolate the effect of selection from other potential sources of variability. 

Price's equation should be viewed as a way to separate the effect of selection from the effect of the environment, rather than as a predictive or quantitative formula to compute them, as remarked in \cite{gardner_prices_2020}. 
The same can be said of all our results, where the strength of selection and the covariance between a trait value and its associated fitness landscape value can be viewed as two possible definitions of selection.
These two notions of natural selection are different because of the distinction between growth rate and fitness landscape: one is concerned by the change in the frequencies of the trait values over time, and is computed by counting individuals, while the other one represents the shift in the frequencies of the trait values at snapshot time $t$ between situations with and without selection, and is based on the comparison between chronological and retrospective samplings of the lineages.

\begin{figure}
	\includegraphics[width=\linewidth]{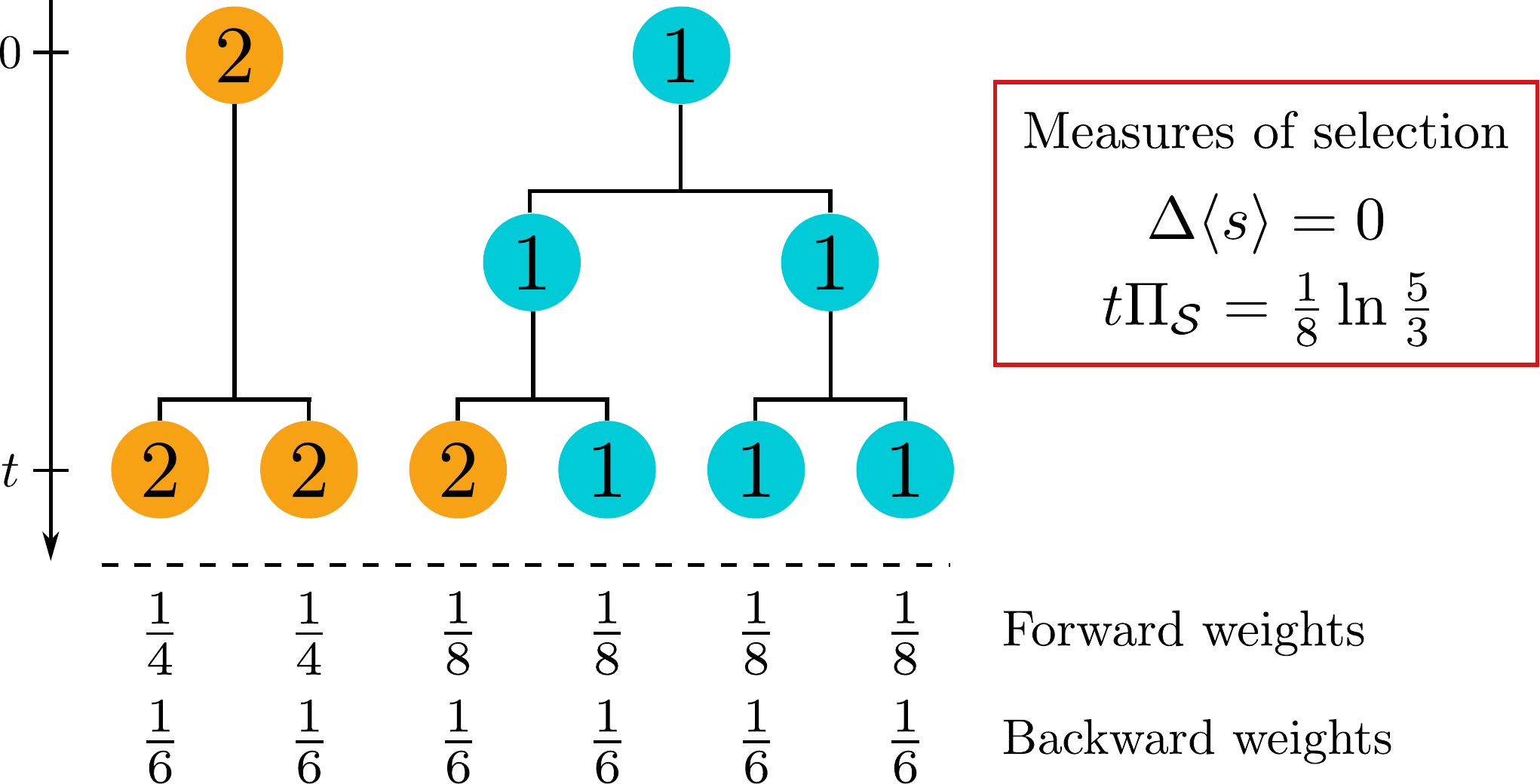}
	\caption{Population tree where cells can have only two values $s=1$ and $s=2$ for the phenotypic trait $\mathcal{S}$. Cells with phenotype $1$ divide more often than cells with phenotype $2$, and phenotypic switching from $1$ to $2$ occurs randomly at division. These two phenomena balance so that the frequencies of $s=1$ and $s=2$ are the same at $t=0$ and at $t$. However, the fitness landscapes for $s=1$ and $s=2$ are different, and are computed using the backward and forward weights of each lineage, leading to $t \Pi_{\mathcal{S}}=\ln \lp 5/3 \rp /8$. The strength of selection is non-zero while the measure of selection proposed by Price, namely $\Delta \langle s \rangle$, is null.}
	\label{fig_price_vs_us}
\end{figure}

Let us give a minimal example for which the strength of selection is non-zero while the mean value of the trait $\mathcal{S}$ is unchanged, because of the balance between heterogeneity in reproductive success and phenotypic switching at division. 
This case is illustrated on \cref{fig_price_vs_us} for a trait $\mathcal{S}$ taking only two values: $s=1$ and $s=2$. Individuals with trait value $1$ reproduce typically twice as fast as those carrying trait value $2$, but they can also switch to trait value $2$ randomly at division. For simplicity, the values $1$ and $2$ of the trait cannot change themselves over time, in other words there is no environment effect here. 
The average value of trait $\mathcal{S}$ is the same at time $t=0$ and at time $t$, the covariance term in Price's equation is zero, and there is no selection in Price's sense. Therefore, from this point of view, there is no difference between this situation and the situation where both values $1$ and $2$ reproduce at the same rate, without phenotypic switching at division. 
One the other hand, individuals with trait value $1$ are over-represented in the backward statistics as compared to the forward statistics, while the opposite is true for trait value $2$, meaning that the fitness landscapes for $s=1$ and $s=2$ are different. Indeed, $p_{\T{back}} (s=1,t) = p_{\T{back}} (s=2,t) =1/2$, $p_{\T{for}} (s=1,t)=3/8$, $p_{\T{for}} (s=2,t)=5/8$, which leads to $t h_t(s=1)=\ln \lp 4/3 \rp + t \Lambda_t$ and $t h_t(s=2)=\ln \lp 4/5 \rp + t \Lambda_t$, with $t \Lambda_t=\ln 3$, using \cref{eq_def_fit_land_s}. 
This difference in fitness landscape results in a non-zero strength of selection, $t \Pi_{\mathcal{S}}=\ln \lp 5/3 \rp /8$, using \cref{eq_def_pi}.
We argue that the strength of selection $\Pi_{\mathcal{S}}$ may be a more appropriate way to define selection, since it gives a non-zero measure of selection for the example discussed above, and thus is more representative of the selection occurring in the population.

\section{A lower bound for the strength of selection}
\label{sec_lower_bbound}

We showed how the equality between the strength of selection and the variance of the fitness landscape distribution, that holds in the Gaussian case, becomes an inequality in general. To complement the upper bound on the strength of selection given by \cref{eq_tur_hs}, we now derive a non-trivial lower bound, which presents an interest to quantify the minimal effect of selection on a particular trait. 

Using a property of the Jeffrey's divergence, the strength of selection can be decomposed as a sum of two Kullback-Leibler (KL) divergences: $\mathcal{J}(p_{\T{back}} | p_{\T{for}}) = \mathcal{D}_{\rm{KL}}(p_{\T{back}} | p_{\T{for}}) + \mathcal{D}_{\rm{KL}}(p_{\T{for}} | p_{\T{back}})$, where $\mathcal{D}_{\rm{KL}}(p | q) = \int p(x) \ln(p(x)/q(x)) \di x$. The positivity of both KL divergences, ensured by Jensen's inequality, gives $\Lambda_t - \langle h_t(s) \rangle_{\T{for}} \geq 0 $ and $\langle h_t(s) \rangle_\T{back} - \Lambda_t  \geq 0$. By combining these two inequalities, we recover that the strength of selection is a positive quantity. 

We can therefore improve the trivial bound on the strength of selection, which is $0$, by improving the two inequalities separately,  using a sharpened version of Jensen's inequality, derived in \cite{liao_sharpening_2019}. Let us now detail how this works in our problem. 

We define the convex functions $\varphi_{\T{for}}(x)=e^{tx}$, $\varphi_{\T{back}}(x)=e^{-tx}$ and the function 
\begin{equation}
\psi(\varphi,x,\nu)=\frac{\varphi(x)-\varphi(\nu)}{(x-\nu)^2}-\frac{\varphi'(\nu)}{x-\nu} \,,
\end{equation}
where $\varphi'$ stands for the derivative of $\varphi$.
The sharpened version of Jensen's inequality reads
\begin{equation}
\langle e^{th_t} \rangle_{\T{for}} -  e^{t \langle h_t \rangle_{\T{for}}} \geq \sigma^2_{\T{for}}(h_t) \ \inf_{h} \psi(\varphi_{\T{for}},h,\langle h_t \rangle_{\T{for}}) \,,
\end{equation}
and is used to improve upon the inequality $\Lambda_t - \langle h_t(s) \rangle_{\T{for}} \geq 0 $. A similar improvement is obtained for $\langle h_t(s) \rangle_\T{back} - \Lambda_t  \geq 0$ by considering $\varphi_{\T{back}}$ instead of $\varphi_{\T{for}}$. 
Combining the two results gives (see \cref{sec_app_low_bound})
\begin{align}
\label{eq_pi_low}
\Pi_{\mathcal{S}} \ \geq \ & \frac{1}{t} \Big[ \frac{\sigma^2_{\T{for}}(h_t)}{\exp(t \Lambda_t)} \ \psi(\varphi_{\T{for}}, h_{\T{min}},\langle h_t \rangle_{\T{for}})  \nn \\
&+ \frac{\sigma^2_{\T{back}}(h_t)}{\exp(-t \Lambda_t)} \ \psi(\varphi_{\T{back}}, h_{\T{max}},\langle h_t \rangle_{\T{back}}) \Big] \,,
\end{align}
which shows that the lower bound depends on the forward and backward variances of the fitness landscape, as well as on its average values and on the minimal (resp. maximal) values of these distributions denoted $h_{\T{min}}$ (resp. $h_{\T{max}}$). 
When the fitness landscape is a monotonic function of the value of the trait, which is the case for cell age and size \cite{genthon_fluctuation_2020}, or for the number of divisions, these extreme values are given by the extreme values of the trait itself. 

Several weaker but simpler forms of this inequality, this time independent of the average fitness landscape values, or independent of the extreme values, or independent of the two, are derived in \cref{sec_app_low_bound}.
In any of those cases, the lower bound is a linear combination of the forward and backward variances of the fitness landscape.

\section{Tests of the linear response relations}
\label{sec_test}

We now illustrate the various bounds for growing cell populations, using both simulations and time-lapse video-microscopy experimental data \cite{kiviet_stochasticity_2014}.

\begin{figure}[t]
	\includegraphics[width=\linewidth]{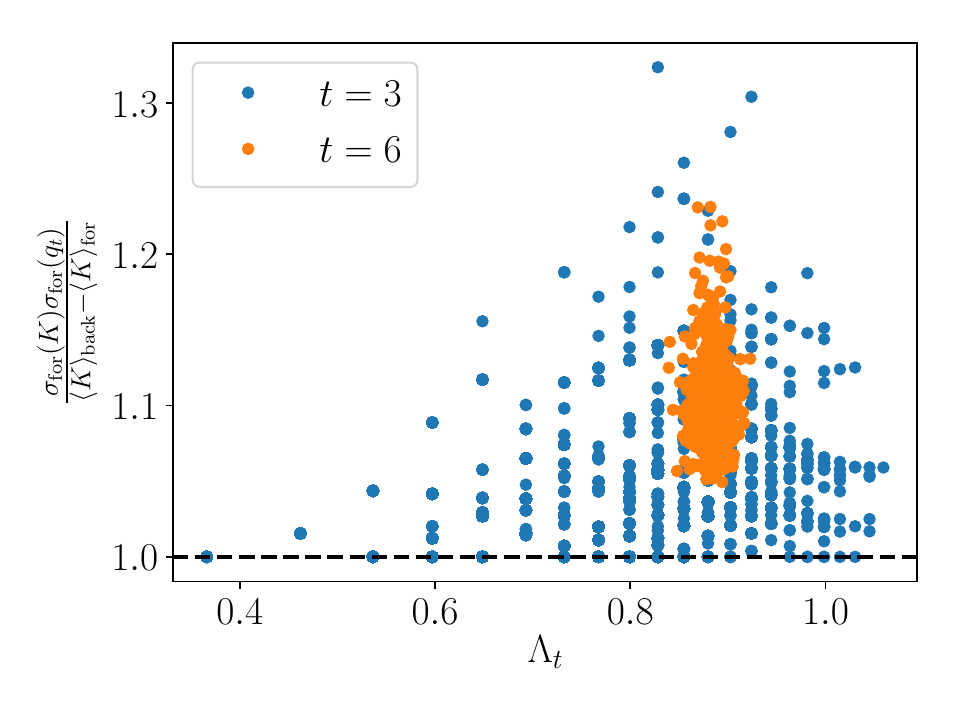}
	\caption{Points of $\sigma_{\T{for}}(K) \sigma_{\T{for}}(q_t)/(\langle K \rangle_{\T{back}}-\langle K \rangle_{\T{for}})$ against $\Lambda_t$ for many tree simulations using a size-controlled model. Each dot corresponds to a single tree, the two sets of data have the same parameters except for the final times of the simulation, which are $t=3$ (blue) and $t=6$ (orange). 
	The black horizontal dashed line at $y=1$ represents the point where the inequality of 
	\cref{eq_tur_hs} is saturated.}
	\label{fig_tur_scat_plot}
\end{figure}

First, we test \cref{eq_tur_s} for the number of divisions $\mathcal{K}$, and for the linear function $g_t(K)=K$, so that the inequality bounds $\langle K \rangle_{\T{back}} - \langle K \rangle_{\T{for}}$.
We simulate lineage trees starting from one cell, for a particular agent-based model in which cells are described by their sizes. Cell sizes continuously increase at constant rate between divisions, and cells divide after a stochastic time only depending on their sizes. Each simulation of such a tree yields a single point on the scatter plot \cref{fig_tur_scat_plot}, which shows the ratio of $\sigma_{\T{for}}(K) \sigma_{\T{for}}(q_t)$ to $\langle K \rangle_{\T{back}}-\langle K \rangle_{\T{for}}$ versus the population growth rate $\Lambda_t$. Two sets of points are presented, which only differ in the final time of the simulation. As expected from \cref{eq_tur_s}, all points in both sets are above $1$.
When the duration of the simulation is small ($t=3$), the final population is small, around $N \sim 20$, therefore for a given tree the lineages do not have time to differentiate significantly and the variability in the number of divisions among the lineages is small. In that case, simulations points are approaching the horizontal dashed line at $y=1$ corresponding to the saturation of the inequality. The final population $N$ fluctuates significantly from one simulation to the next, because the simulation time is short and all simulations start with a single cell with random initial size. As a result, the dispersion of values of $\Lambda_t$ is large. 

Now, when doubling the duration of the simulation, the cloud of scattered points is considerably reduced in both directions. The horizontal dispersion reduces because as $t$ increases, the state of the system at the final time becomes less and less affected by the initial condition. On the vertical axis, there is a gap between the lower part of the scatter plot and the horizontal line at $y=1$ due to the increase of heterogeneity in the number of divisions in the lineages with the simulation time. 
\begin{figure*}
	\includegraphics[width=\linewidth]{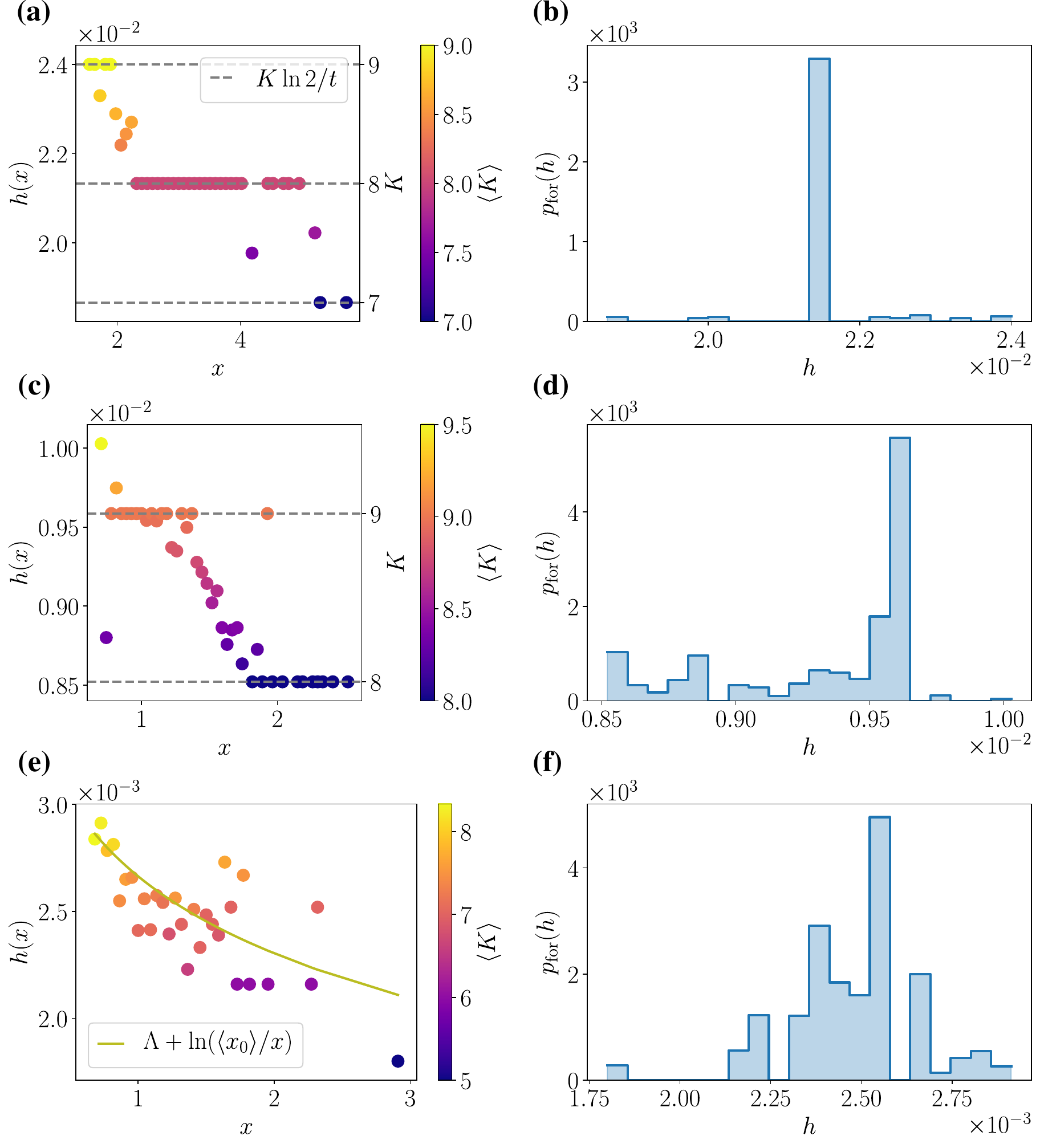}
	\caption{Experimental fitness landscapes for size and their forward distributions, computed with data from \cite{kiviet_stochasticity_2014}. Each row of the figure corresponds to a different experiment and the first column shows fitness landscapes $h_t(x)$ as functions of size $x$.
	\textbf{(a,c)}: the grey horizontal dashed lines correspond to theoretical plateaus, equal to $K \ln 2 /t$, predicted  when $K$ is fully determined by the value $s$ of the trait. The integers $K$ corresponding to the plateaus are indicated on the right $y$-axis.
	\textbf{(e)}: plateaus are blurred and replaced by a smoother scatter plot in good agreement with the general shape of the theoretical prediction, made in the case where there is no variability in individual growth rate nor volume partition at division \cite{genthon_fluctuation_2020}. $\Lambda$ is the population growth rate and $\langle x_0 \rangle$ the average size of initial cells.
	\textbf{(a,c,e)}: each dot is made of all the cells having the same size, and the mean number of divisions amongst those cells is represented by the color of the dot. This shows that dots aligning on a plateau corresponding to a number $K$ of divisions truly come from cells that underwent $K$ divisions.
	\textbf{(b,d,f)}: the second column represents the distribution $p_{\rm{for}}(h)$ of the corresponding size fitness landscapes (i.e. on the same row) with the forward size distribution. For \textbf{(b,d)}, the fitness landscapes are highly non-Gaussian, and the peaks in these distributions correspond to the value of one of the plateaus.}
	\label{fig_hx}
\end{figure*}

Second, we test our results on real data extracted from \cite{kiviet_stochasticity_2014}, which are made of $11$ population trees corresponding to the growth of E. Coli in different nutrients.
We focus on the number of divisions $\mathcal{K}$, the size $\mathcal{X}$ and the age $\mathcal{A}$, which are easily accessible and for which we previously studied the theoretical fitness landscapes \cite{genthon_fluctuation_2020}. 
Only plots for the size are presented in the main text, and similar plots for the age can be found in \cref{sec_app_h_age}.

The first step is to determine the fitness landscapes, which are shown for three particular experimental conditions on \cref{fig_hx}. Each row corresponds to a particular experiment, the first column displays the fitness landscapes as functions of the size $x$, and the second column shows the distributions of the corresponding fitness landscapes, computed with the forward size distributions. 
It is straight-forward to demonstrate that in the case where the number of divisions $K$ is completely determined by the value $s$ of the trait, then $h_t(s)=h_t(K(s))=K(s) \ln 2/t$ \cite{nozoe_inferring_2017,genthon_fluctuation_2020}. In this case, the fitness landscape is an ensemble of plateaus corresponding to the values of $K$ featured in the population at time $t$, and cells on the same plateau have undergone the same number of divisions, even though they have a different value $s$. 
Those predicted plateaus are actually observed for several experiments, as shown on \cref{fig_hx}a,c. Indeed, we evaluate the mean number of divisions for the set of cells used to compute each point of the fitness landscape, and represent it with a color code. We also plot the theoretical plateaus, with the discrete number of divisions $K$ corresponding to the plateau on the right $y$-axis of each plot. We see a very good agreement between the mean number of divisions of cells aligned on a particular plateau and the value $K$ corresponding to this plateau on the $y$-axis. This suggests a strong correlation between the value of the size and the number of divisions on the lineage.	
The dots between the plateaus correspond to sizes that have been reached by cells with different numbers of divisions (leading to non-integer mean values), as highlighted by the gradation from one color to another. 
By going from the top experiment to the bottom one, the plateaus gradually blur and are replaced on \cref{fig_hx}e by a smoother curve, in good agreement with the logarithmic prediction we made in \cite{genthon_fluctuation_2020}. This happens when lineages de-synchronize because of the cumulative effect of various noises, leading to a weaker dependence of the number of divisions $K$ on the final value of the trait $s$. 

On the right column of the figure, we see that fitness landscapes strongly deviate from being normally distributed, which justifies the need to go beyond the results known in the Gaussian case. More precisely, on \cref{fig_hx}b,d, fitness landscapes exhibit peaks at values of $h$ corresponding to one of the plateaus appearing on the left-column plot. We notice that not all the 3 plateaus of \cref{fig_hx}a (resp. 2 plateaus of \cref{fig_hx}b) are mapped with a peak in the corresponding forward fitness landscape distribution on \cref{fig_hx}b (resp. \cref{fig_hx}d). This is because the cell size distribution tend to zero for extreme sizes (that is $0$ and $+ \infty$), thus cells of large sizes aligning on plateaus defined by small $K$ and cells of small sizes aligning on plateaus defined by large $K$ contribute very little to the cell size distribution and thus to the fitness landscape distribution.
\begin{figure}[t]
	\includegraphics[width=\linewidth]{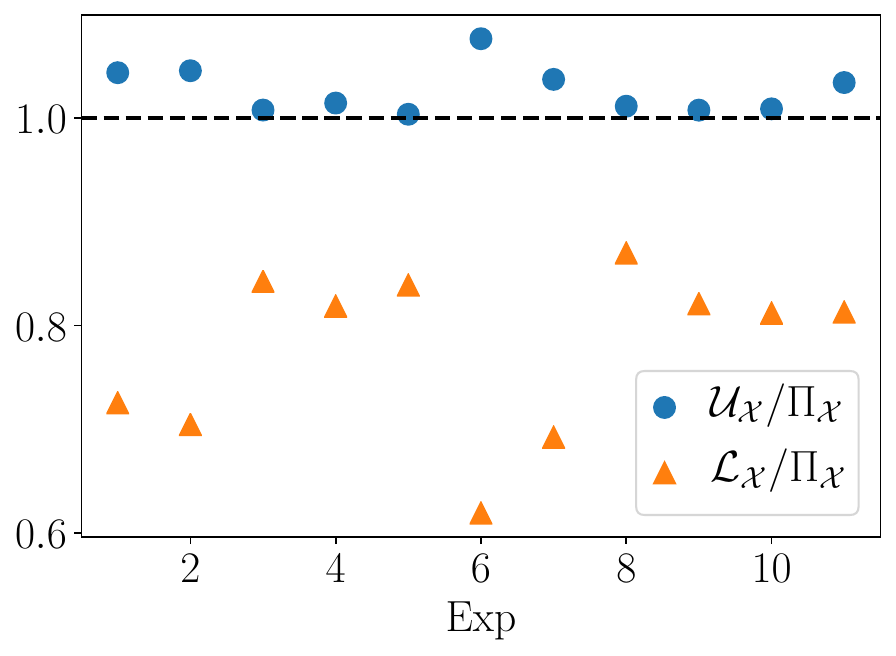}
	\caption{Upper bound $\mathcal{U}_{\mathcal{X}}$ (blue dots) and lower bound $\mathcal{L}_{\mathcal{X}}$ (orange triangles) for the strength of selection acting on size $\Pi_{\mathcal{X}}$, normalized by the latter. The $x$-axis represents the $11$ colonies in different growth conditions from \cite{kiviet_stochasticity_2014}, in no particular order. }
	\label{fig_pix_bounds}
\end{figure}

Then, we test the upper and lower bounds on the strength of selection acting on cell size using these data. 
We show on \cref{fig_pix_bounds} the upper bound $\mathcal{U}_{\mathcal{X}}$ given by \cref{eq_tur_hs} and the lower bound $\mathcal{L}_{\mathcal{X}}$ given by \cref{eq_pi_low}, normalized by the strength of selection $\Pi_{\mathcal{X}}$. The $x$-axis labels in no particular order the colonies which have grown in different nutrient medium \cite{kiviet_stochasticity_2014}. As expected, points representing the upper bound and those representing the lower bound are respectively above and below the horizontal dashed line at $y=1$. 
Experiments for which the normalized upper bound approaches $1$ indicate that cell cycles are almost synchronized and thus that there is small variability in terms of number of divisions among the lineages. 

Note that Nozoe et al. also proved \cite{nozoe_inferring_2017} that the strength of selection for the division bounds the strength of selection acting on any trait: $0 \leq \Pi_{\mathcal{S}} \leq \Pi_{\mathcal{K}}$. This bound, is typically not as tight as \cref{eq_tur_hs} (see \cref{sec_app_h_age} for comparison). To improve upon it, one can use $\mathcal{S}=\mathcal{K}$ in \cref{eq_tur_hs} to obtain a bound for $\Pi_{\mathcal{K}}$ itself.

\section{Discussion}
\label{sec_discussion}

The general idea of comparing the response of a system in the presence of a perturbation 
to its fluctuations in the absence of the perturbation lies at the heart of the Fluctuation-Dissipation theorem, which has a long history of physics, with some applications to evolution \cite{leibler_individual_2010,kaneko_macroscopic_2018}. Remarkably, the present framework with forward (unperturbed) and backward (perturbed) dynamics can be conveniently applied to population dynamics without having to perform additional experiments, since both probabilities can be calculated with the same lineage tree. Our main result is a set of inequalities for the average of an arbitrary function of a trait or for its fitness landscape, valid beyond the Gaussian assumption, and which constrain the strength of selection in population dynamics even in the presence of time-dependent selection pressures.

These inequalities are universal because they only rely on the branching structure of the population tree and are completely independent of the dynamics of the tree, that is the ensemble of rules governing the division of the branches.
In the context of cell populations, this means for instance that our results are valid for any control strategy (sizer, timer, adder, ...) and in presence of any source of noise (variability in single cell growth rate or size at division, asymmetry in resource partitioning between sister cells, ...) \cite{lin_effects_2017}, in presence of possible mutations, and regardless of the nature of the cell (bacteria, yeast, stem-cell, ...).
Although we illustrated our results with cell populations, they apply to any stochastic process defined on a branched tree. In particular, they could be insightful in the context of ecology, where such trees are used to represent phylogeny \cite{schuh_biological_2009}. In this case, each lineage could represent a species or a version of a gene, instead of an individual, and the divisions would represent speciations or mutations, respectively. The notions of fitness landscape and selection strength appear meaningful in this setting, as a quantification of the correlations between a feature of a species/genetic information and the number of speciations/mutations its phylogenetic lineage underwent. 

For applications, we focused on phenotypic traits, like cell size or age, and in these cases, we found our 
upper bound on the strength of selection to be tight. In future work, it would be interesting to apply this framework to genotypic traits instead of phenotypic ones \cite{neher_statistical_2011} and possibly exploit recent methods of lineage tracking \cite{nguyen_ba_high-resolution_2019}. This could open new perspectives to address a number of important problems like antibiotic resistance \cite{lambert_quantifying_2015}, the differentiation of stem cells \cite{tak_simultaneous_2019} or virus evolution.

The search for universal principles in evolution is an active field of research \cite{goldenfeld_life_2011,kaneko_macroscopic_2018}. An important step in this endeavor was made by Fisher, who boldly compared his theorem to the second law of thermodynamics. While the theorem turned out not to be as general as expected, Fisher had nevertheless the correct intuition about its importance for evolutionary biology, and he was also correct in expecting that such a general principle should be related to thermodynamics.

\begin{acknowledgments}
	DL acknowledges insightful discussions with L. Peliti, O. Rivoire, J. Unterberger and C. Loverdo and support from Agence Nationale de la Recherche (ANR-10-IDEX-0001-02, IRIS OCAV and (ANR-11-LABX-0038, ANR-10-IDEX-0001-02).
\end{acknowledgments}


\appendix

\section{Phenotypic fitness landscape and growth rate associated with the value $s$ of a trait $\mathcal{S}$}
\label{sec_app_h_vs_lambda}

The fitness landscape $h_t(s)$ of trait $\mathcal{S}$ defined in \cref{eq_def_fit_land_s} should not be confused with the fitness associated with the value $s$ of the trait, which can be identified with the growth rate of the subpopulation carrying that trait. However, they are linked by a simple relation. In line with the definition of the population growth rate, we define the growth rate of the subpopulation carrying the value $s$ of the trait $\mathcal{S}$ as
\be
\label{eq_app_lams_lam}
\Lambda_t(s)=\frac{1}{t} \ln \lbk \frac{n(s,t)}{n(s,0)} \rbk = \Lambda_t + \frac{1}{t} \ln \lbk \frac{p_{\T{back}} (s,t)}{p_{\T{back}} (s,0)} \rbk \,,
\ee
where $n(s,t)= p_{\T{back}} (s,t) N(t)$ is the number of cells with trait value $s$ at time $t$. Of course, defining this quantity only makes sense between two times $t=0$ and $t$ for which the value $s$ is present in the population, otherwise the expression in undefined. 

By comparing \cref{eq_def_fit_land_s,eq_app_lams_lam}, we can link the growth rate associated with a trait value $s$ to the fitness landscape of this trait value:
\be
\label{eq_app_hs_lams}
h_t(s)- \Lambda_t(s) = \frac{1}{t} \ln \lbk \frac{p_{\T{back}}(s,0)}{p_{\T{for}}(s,t)}  \rbk = \frac{1}{t} \ln \lbk \frac{p_{\T{for}}(s,0)}{p_{\T{for}}(s,t)} \rbk \,. 
\ee
The last equality follows from the fact that, at $t=0$, the cells have not divided yet and so the forward and backward samplings of the population are identical. 

Finally, we see that the three quantities $\Lambda_t$, $\Lambda_t(s)$ and $h_t(s)$ are different but intimately linked. In the long time limit, for which equilibrium distributions do not depend on time anymore, they all become equal to the steady-state population growth rate $\Lambda$
\be
\underset{t \rightarrow \infty}{\lim} \Lambda_t= \underset{t \rightarrow \infty}{\lim} \Lambda_t(s) = \underset{t \rightarrow \infty}{\lim} h_t(s)= \Lambda \,.
\ee
Moreover, as we already noted, \cref{eq_def_fit_land_s} indicates that the sign of 
$h_t(s)-\Lambda_t$ informs us on the comparison between the backward and forward probabilities of trait value $s$ in the population at time $t$, or in other words, if the trait value $s$ is over-represented in the population as compared to a situation without selection. Thus, $h_t(s)-\Lambda_t$ quantifies the separate effect of selection.

Following the intuitive understanding of the growth rate associated with a trait value $s$, \cref{eq_app_lams_lam} means that the sign of $\Lambda_t(s)-\Lambda_t $ informs us on the comparison between the backward statistics of the trait value $s$ at time $t$ and at time $0$. A trait value is favored by the population dynamics, which includes all phenomena leading to changes in trait value frequencies, if its growth rate is larger than the population growth rate, which corresponds to an increase of the frequency of that trait value in the population as time grows.

The last relation, \cref{eq_app_hs_lams} provides a new insight: the sign of $h_t(s)-\Lambda_t(s)$ is linked to the comparison between the forward probability of trait value $s$ at time $0$ and time $t$. The forward statistics is constructed to balance the effect of selection occurring in tree-structured data. However, it is affected by all the other sources of variability, as for example mutations. Therefore, the sign of $h_t(s)-\Lambda_t(s)$ indicates the evolution of the frequency of the value $s$ of trait $\mathcal{S}$ as time grows, due to every phenomenon but selection. 

The \cref{eq_def_fit_land_s} can be written in the form of a fluctuation relation \cite{nozoe_inferring_2017,genthon_fluctuation_2020}, with an exponential bias between the forward and backward probabilities, similar to Crooks fluctuation relation for work in stochastic thermodynamics.
The same can be done with \cref{eq_app_lams_lam,eq_app_hs_lams}, however, unlike \cref{eq_def_fit_land_s}, these new fluctuation relations both link two probability distributions that may not have the same support.

\section{Gaussian case}
\label{sec_app_gauss}

We show in this section that a linear relation between the strength of selection and the variance of the fitness landscape holds in the case where the fitness landscape is normally distributed. 
To do so, let us first derive a very useful result: by isolating $p_{\T{back}}(s,t)$ in \cref{eq_def_fit_land_s}, and integrating over $s$ using the normalization of $p_{\T{back}}$, the population growth rate is expressed as a forward average
\begin{equation}
e^{t \Lambda_t} = \ \langle e^{t h_t} \rangle_{\T{for}} \,.
\label{eq_app_lambda_for}
\end{equation}
We can do the same but isolating $p_{\T{for}}(s,t)$ this time, leading to 
\begin{equation}
e^{-t \Lambda_t} = \ \langle e^{-t h_t} \rangle_{\T{back}} \,.
\label{eq_app_lambda_back}
\end{equation}

We now assume that $h_t(s)$ can be accounted for by a continuous probability distribution, even though the trait $\mathcal{S}$ may be discrete, as it in the case for the number of divisions. 
We set a Gaussian forward distribution with mean $\langle h_t \rangle_{\T{for}}$ and variance $\sigma_{\T{for}}(h)^2$ for the fitness landscape $h_t(s)$, then $ \exp(t h_t(s))$ follows a log-normal distribution of mean
\be
\label{eq_app_lamb_gaus}
\langle e^{t h_t} \rangle_{\T{for}} = e^{t \langle h_t \rangle_{\T{for}} + (t \sigma_{\T{for}}(h_t))^2/2} \,.
\ee
This relation shows that for a given forward average fitness landscape, the growth rate is positively affected by the variability between the lineages.

The backward average of the fitness landscape is given by the forward average of a biased fitness landscape:
\begin{equation}
\langle h_t \rangle_{\T{back}} = e^{-t \Lambda_t} \int h_t(s) e^{th_t(s)} p_{\T{for}} (s,t) \di s
\end{equation}
We make the hypothesis that the fitness landscape is a bijective function of the trait value and use the conservation of the probability: $p_{\T{for}} (s,t) \di s = p_{\T{for}} (h) \di h$, leading to a solvable Gaussian integral
\begin{align}
\label{eq_app_h_back_gaus}
&\langle h_t \rangle_{\T{back}}  \nn \\
& = \frac{e^{-t \Lambda_t}}{\sqrt{2 \pi \sigma_{\T{for}}(h_t)^2}} \int h e^{th} e^{-(h-\langle h_t \rangle_{\T{for}})^2 /(2 \sigma_{\T{for}}(h_t)^2) } \di h \nn \\
& = e^{-t \Lambda_t} \lp \langle h_t \rangle_{\T{for}} + t \sigma_{\T{for}}(h_t)^2 \rp e^{t \langle h_t \rangle_{\T{for}} + (t \sigma_{\T{for}}(h_t))^2/2} \,.
\end{align}
Finally, combining \cref{eq_app_lambda_for,eq_app_lamb_gaus,eq_app_h_back_gaus}, we obtain
\be
\label{eq_back_for_gauss}
\langle h_t \rangle_{\T{back}}= \langle h_t \rangle_{\T{for}} + t \sigma_{\T{for}}(h_t)^2 \,,
\ee
and thus
\be
\label{eq_app_pis}
\Pi_{\mathcal{S}} = t \sigma_{\T{for}}(h_t)^2 \,.
\ee
%

Moreover, combining \cref{eq_app_lamb_gaus,eq_app_pis} we deduce that $\langle h_t \rangle_{\T{for}}$ and $\langle h_t \rangle_{\T{back}}$ are not only respectively smaller and greater than $\Lambda_t$, as discussed in \cref{sec_lower_bbound}, but they are actually symmetrical around this value $\langle h_t \rangle_{\T{back}}- \Lambda_t = \Lambda_t - \langle h_t \rangle_{\T{for}} = t \ \sigma^2_{\T{for}}(h_t)/2$.
In other words, in this particular case, the KL divergence is symmetrical: $\mathcal{D}_{\rm{KL}}(p_{\T{for}} | p_{\T{back}})=\mathcal{D}_{\rm{KL}}(p_{\T{back}} | p_{\T{for}})$.

In the case where $h_t(s)$ follows a Gaussian distribution in the forward statistics, it also follows a Gaussian distribution in the backward statistics because the bias of the fluctuation relation between $p_{\T{back}}$ and $p_{\T{for}}$ is exponential in $h_t$. Since $h_t$ follows a Gaussian distribution of mean $\langle h_t \rangle_{\T{back}}$ and standard deviation $\sigma_{\T{back}}(h_t)$ in the backward statistics, then $ \exp \lbk -t h_t(s) \rbk$ follows a log-normal distribution of mean
\be
\langle e^{-t h_t} \rangle_{\T{back}} =  e^{-t \langle h_t \rangle_{\T{back}} + (t \sigma_{\T{back}}(h_t))^2/2} \,.
\ee
We now take the inverse of this formula and use \cref{eq_app_lambda_back,eq_back_for_gauss} to replace the backward average:
\be
\label{eq_app_lamb_gaus_back}
e^{t \Lambda_t}= e^{t \lp \langle h_t \rangle_{\T{for}} + t \sigma_{\T{for}}(h_t)^2 \rp - (t \sigma_{\T{back}}(h_t))^2/2} \,.
\ee
By comparing \cref{eq_app_lamb_gaus,eq_app_lamb_gaus_back}, it follows that $\sigma_{\T{back}}(h_t)=\sigma_{\T{for}}(h_t)$. 
Finally, the standard deviation in \cref{eq_app_pis} can be taken indifferently with respect to both statistics and we omit the index to write the general version of \cref{eq_app_pis}:
\be
\Pi_{\mathcal{S}} = t \mathrm{Var}(h) \,.
\ee

\section{Upper bounds numerical comparison}
\label{sec_app_bounds_vs_sasa}

\begin{figure*}[t]
	\includegraphics[width=\linewidth]{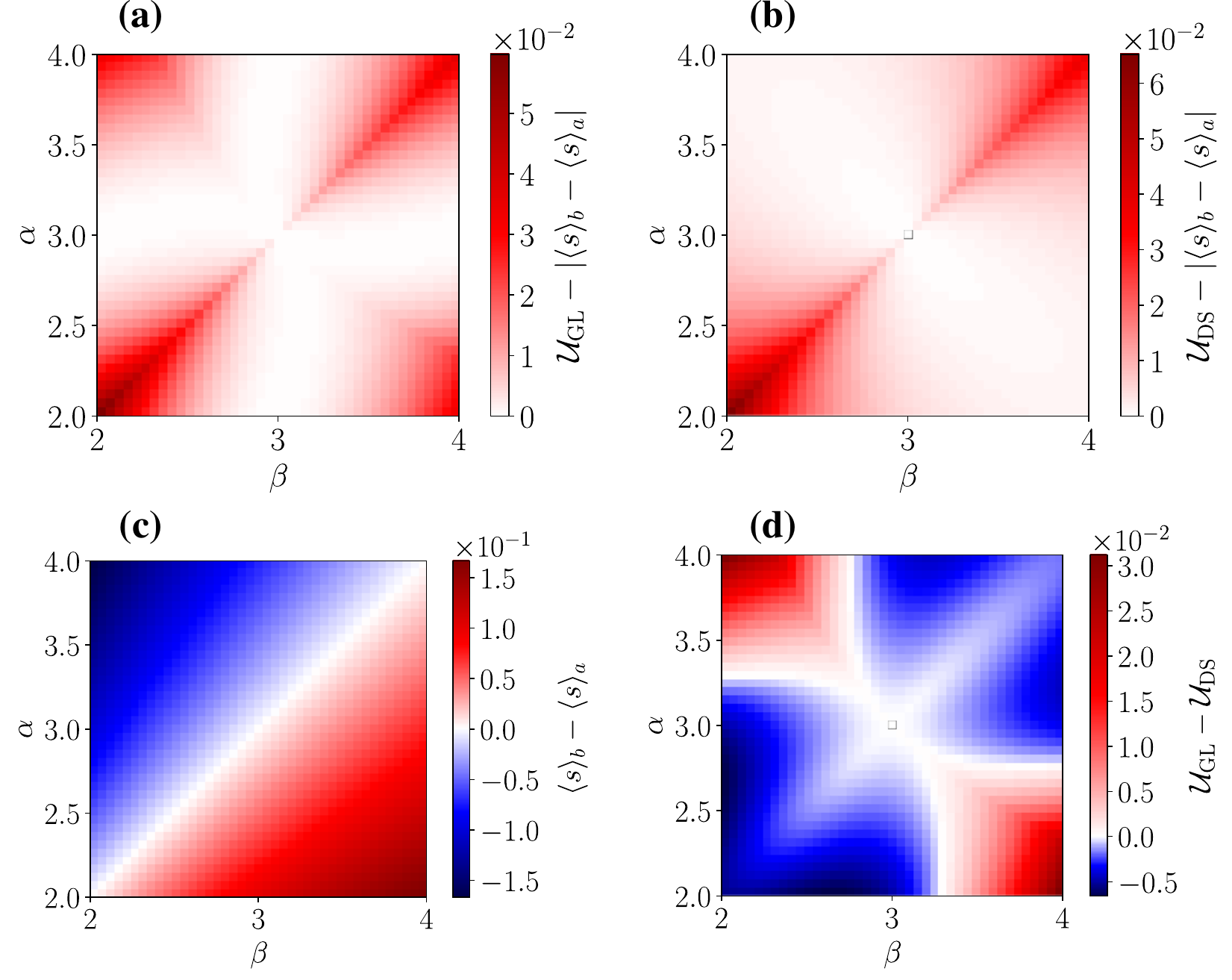}
	\caption{Comparison between the upper bounds $\mathcal{U}_{\rm{GL}}$ (\cref{eq_app_def_ugl}) and $\mathcal{U}_{\rm{DS}}$ derived in \cite{dechant_fluctuationresponse_2020} (\cref{eq_app_def_uds}), for beta distributions $p_a(s)=f(s,\alpha,\beta)$ and $p_b(s)=f(s,3,3)$. Parameters $\alpha$ and $\beta$ are varied between $2$ and $4$. 
	First row: difference between the upper bounds and $| \langle s \rangle_{\T{b}}-\langle s \rangle_{\T{a}}|$, \textbf{(a)} for our bound, and \textbf{(b)} for Dechant-Sasa's bound, showing that all points are indeed above $0$. 
	\textbf{(c)}: exact difference $\langle s \rangle_{\T{b}}-\langle s \rangle_{\T{a}}$, in agreement with the theoretical value $\langle s \rangle_{\T{b}}-\langle s \rangle_{\T{a}} = 1/2 - \alpha / (\alpha + \beta)$.
	\textbf{(d)}: comparison between $\mathcal{U}_{\rm{GL}}$ and $\mathcal{U}_{\rm{DS}}$, blue regions indicate where our bound is tighter, i.e. smaller, and red regions indicate where Dechant-Sasa's bound is tighter.	
	For all four plots, the grid is $41 \times 41$, and numerical values are rounded to $10^{-15}$ to avoid python floats precision errors.}
	\label{fig_app_bounds_vs_sasa}
\end{figure*}

In this section we compare numerically the upper bound $\mathcal{U}_{\rm{GL}}$ obtained in \cref{eq_tur_s_min}:
\begin{equation}
\label{eq_app_def_ugl}
\mathcal{U}_{\rm{GL}}=\min  \lp \sigma_{\T{a}}(g_t) \sigma_{\T{a}}(q_t), \sigma_{\T{b}}(g_t) \sigma_{\T{b}}(r_t) \rp \,,
\end{equation}
to the upper bound $\mathcal{U}_{\rm{DS}}$ derived by Dechant and Sasa (Eq. $5$ in \cite{dechant_fluctuationresponse_2020}):
\begin{equation}
\label{eq_app_def_uds}
\mathcal{U}_{\rm{DS}} = \underset{\gamma > 0}{\rm{inf}} \lp K_{g_t}^a (\gamma \sigma) -\gamma \sigma \langle g_t \rangle_{\rm{a}} + \mathcal{D}_{\rm{KL}}(p_{\T{b}} | p_{\T{a}}) \rp \,,
\end{equation}
where $\sigma={\rm{sign}} (\langle g_t \rangle_{\rm{b}}-\langle g_t \rangle_{\rm{a}})$ and $K_{g_t}^a (\gamma)= \ln \langle \exp \lp \gamma g_t \rp \rangle_{\rm{a}}$ the cumulant-generating function of $g_t$.	
Both quantities $\mathcal{U}_{\rm{GL}}$ and $\mathcal{U}_{\rm{DS}}$ bound the difference $| \langle g_t \rangle_{\T{b}}-\langle g_t \rangle_{\T{a}}|$ between the average values of a function $g_t$ of a variable $\mathcal{S}$ with respect to probability distributions $p_a$ and $p_b$.

To compare them, we took beta distributions for $p_a$ and $p_b$, having the same support $[0,1]$ so that both bounds are defined. Beta-distributed variables admit a probability density function (pdf) of the form $f(s,\alpha,\beta)=B(\alpha,\beta) s^{\alpha-1} (1-s)^{\beta-1}$, where $B(\alpha,\beta)$ is a normalization constant, and their mean is given by $\langle s \rangle = \alpha / (\alpha + \beta)$. 	
We fix the pdf in the ensemble $b$ to $p_b(s)=f(s,3,3)$, whose bell shape is reminiscent of a Gaussian distribution, on a finite interval. 
The pdf in the ensemble $a$ is taken as $p_a(s)=f(s,\alpha,\beta)$, where $\alpha$ and $\beta$ are varied in $[2,4]$. 
We choose the simple function $g_t(s)=s$.

Results are shown on \cref{fig_app_bounds_vs_sasa}. The first row of figures shows the difference between the upper bound and the actual difference $| \langle s \rangle_{\T{b}}-\langle s \rangle_{\T{a}}|$, for our bound $\mathcal{U}_{\rm{GL}}$ (\cref{fig_app_bounds_vs_sasa}a), and for Dechant-Sasa's bound $\mathcal{U}_{\rm{DS}}$ (\cref{fig_app_bounds_vs_sasa}b). As expected, all points on these two plots are positive.	
We plot on \cref{fig_app_bounds_vs_sasa}c the real difference $\langle s \rangle_{\T{b}}-\langle s \rangle_{\T{a}}$, which is in complete agreement with the theory: $\langle s \rangle_{\T{b}}-\langle s \rangle_{\T{a}} = 1/2 - \alpha / (\alpha + \beta)$.	
Finally, \cref{fig_app_bounds_vs_sasa}d shows a comparison between our bound and Dechant-Sasa's bound: blue regions represent sets of parameters $(\alpha, \beta)$ where our bound is numerically tighter, and the opposite is true in red regions. 
We note that, if the blue region is larger than the red region, on the other hand the advantage of one bound over the other $  |\mathcal{U}_{\rm{GL}}-\mathcal{U}_{\rm{DS}} |$, is generally larger in the red region.	
Therefore, the answer to the question `Which bound is tighter?' depends on the actual distributions $p_a(s)$ and $p_b(s)$. However, we note that our bound is easier to compute since it does not require the optimization over an external parameter, which is the case for $\mathcal{U}_{\rm{DS}}$ in \cite{dechant_fluctuationresponse_2020} (parameter $\gamma$ in \cref{eq_app_def_uds}). Note that this optimization can be bypassed by choosing a specific value $\gamma$ in \cref{eq_app_def_uds}, but then the corresponding bound is less tight than the version with the infimum.

\section{Information-theoretic-distance $\sigma(q)$ between perturbed and unperturbed dynamics in terms of measurable quantities}
\label{sec_app_std_q}

We show in this section how the information-theoretic-distance $\sigma_{\T{for}}(q_t)$ is linked to measurable quantities in cell colonies, for a general trait $\mathcal{S}$, and for the particular case of age models. 
Combining \cref{eq_def_fit_land_s,eq_def_q} give the ratio $q_t(s)=\exp \lbk h_t(s)-\Lambda_t \rbk$, which combined with \cref{eq_app_lambda_for} lead to \cref{eq_std_q} in the main text.
Thus, $\sigma_{\T{for}}(q_t)$ is the coefficient of variation in the forward statistics of the quantity 
\begin{align}
e^{th_t(s)}&=e^{t \Lambda_t} \frac{p_{\T{back}} (s,t)}{p_{\T{for}} (s,t)} \nn \\
& = \frac{N(t)}{N_0}\frac{p_{\T{back}} (s,t)}{p_{\T{for}} (s,t)} \nn \\
& = \frac{n(s,t)}{N_0}\frac{1}{p_{\T{for}} (s,t)} \,,
\end{align}
where $n(s,t)=N(t) p_{\T{back}} (s,t)$ is the number of cells with the value $s$ of trait $\mathcal{S}$ at time $t$.

In the case where $\mathcal{S}$ is the number of divisions, the fitness landscape is called the lineage fitness and is given by $h_t(K)=K \ln m /t$. Therefore, $\sigma_{\T{for}}(q_t)=\sigma_{\T{for}}(m^K)/\langle m^K  \rangle_{\T{for}}$ is the relative fluctuation of quantity $m^K$, representing the expected number of lineages that underwent $K$ divisions, normalized by the initial population $N_0$, divided by the forward probability of $K$.

For age models, where the division is only controlled by the age of the cell, we know a fluctuation relation linking the forward and backward distributions of generation times $\tau$, defined as the time between two consecutive divisions on the same lineage \cite{powell_growth_1956}
\begin{equation}
\label{eq_fr_tau}
f_{\T{back}}(\tau) =m f_{\T{for}}(\tau) \exp[-\Lambda \tau] \,.
\end{equation}
This relation can be understood as a version of the fluctuation relation on the number of divisions \cite{nozoe_inferring_2017,genthon_fluctuation_2020} at the scale of the cell cycle. However, let us note two differences: first, unlike the one for the number of divisions, \cref{eq_fr_tau} is only true in the long time limit, when the population is growing at a constant steady state growth rate $\Lambda$; and second, the distributions $f_{\T{back}}$ and $f_{\T{for}}$ are not snapshot distributions at time $t$, but distributions of generation times computed along the weighted lineages. 

We define $q(\tau)=f_{\T{back}}(\tau)/f_{\T{for}}(\tau)$ in the same way, and following the same steps for a general function $g(\tau)$ we derive 
\begin{multline}
| \langle g(\tau) \rangle_{\T{back}}-\langle g(\tau) \rangle_{\T{for}} | \leq  \\
\min \lp \sigma_{\T{for}}(g(\tau)) \sigma_{\T{for}}(q(\tau)), \sigma_{\T{back}}(g(\tau)) \sigma_{\T{back}}(r(\tau)) \rp \,.
\end{multline}
From \cref{eq_fr_tau}, we express the information-theoretic-distance term as
\begin{equation}
\sigma_{\T{for}}(q(\tau))=\frac{\sigma_{\T{for}}(e^{-\Lambda \tau})}{\langle e^{-\Lambda \tau} \rangle_{\T{for}}} \,,
\end{equation}
which is the relative fluctuation in the forward sampling of the quantity $\exp \lbk -\Lambda \tau \rbk=f_{\T{back}}(\tau)/m f_{\T{for}}(\tau)$. We know from \cite{hashimoto_noise-driven_2016} that the backward distribution is also the generation time distribution of the direct ancestor cells. Therefore, $\exp \lbk -\Lambda \tau \rbk$ represents the ratio of the probability for the ancestor cell to divide at age $\tau$ to the expected number of daughter cells born from that division that divide at age $\tau$.

\begin{widetext}
	
\section{Small variability limit}
\label{sec_app_small_var}

In this section, we study the two sides of the fluctuation-response inequality on an arbitrary function $g_t$ of a phenotypic trait $\mathcal{S}$ (\cref{eq_tur_s_min}) in the limit where the forward and backward distributions approach each other, and show that they are mathematically equivalent in this limit in the case where the function $g_t$ is the fitness landscape $h_t$.
The difference between the two distributions is captured by the `distance' term $\sigma(q_t(s))$, or equivalently $\sigma(\ln q_t(s))= \sigma (\ln (p_{\T{back}}(s,t)/p_{\T{for}}(s,t))) = t \sigma(h_t(s))$, where the standard deviations can be taken either in the backward or forward statistics. 
From now on, we refer to the limit where the forward and backward distributions are close to each other as the small variability limit, defined by $t \sigma(h_t(s)) \to 0$. 

We first use this limit in the forward statistics: $t \sigma_{\T{for}}(h_t(s)) \to 0$. Starting from \cref{eq_def_fit_land_s}, we isolate $p_{\T{back}}(s,t)$, multiply both sides by $g_t(s)$ and integrate over $s$, leading to the expression of the backward average of function $g_t$ as the forward average of a biased version of the same function:
\be
\label{h_back}
\langle g_t \rangle_{\T{back}} = e^{t (\langle h_t \rangle_{\T{for}}-\Lambda_t)} \int g_t(s) e^{t (h_t(s)-\langle h_t \rangle_{\T{for}})} p_{\T{for}} (s,t) \di s \,.
\ee

In order to expand the exponential, we assume that for any $s$, $t (h_t(s) -\langle h_t \rangle_{\T{for}})$ is small, which corresponds to $t \sigma_{\T{for}}(h_t)$ small because $\sigma_{\T{for}}(h_t)$ is the characteristic distance to the mean. Therefore,
\begin{align}
\label{eq_app_h_back}
\langle g_t \rangle_{\T{back}} & \underset{t \sigma \rightarrow 0}{\sim} e^{t (\langle h_t \rangle_{\T{for}}-\Lambda_t)} \int g_t(s) \lp 1+ t (h_t(s)-\langle h_t \rangle_{\T{for}}) \rp  p_{\T{for}} (s,t) \di s \nn \\
& \sim e^{t (\langle h_t \rangle_{\T{for}}-\Lambda_t)} \lbk \langle g_t \rangle_{\T{for}} + t (\langle g_t h_t \rangle_{\T{for}}- \langle g_t \rangle_{\T{for}}\langle h_t \rangle_{\T{for}}) \rbk \,,
\end{align}
where $\langle g_t h_t \rangle_{\T{for}}- \langle g_t \rangle_{\T{for}}\langle h_t \rangle_{\T{for}}= \mathrm{Cov_{\T{for}}} (h_t,g_t)$ is the covariance of $h_t$ and $g_t$ with respect to the forward probability.
The term in the bracket is a first-order correction to $\langle g_t \rangle_{\T{for}}$ in $t \sigma_{\T{for}}(h_t)$. Now we need to compute the prefactor $\exp[-t \Lambda_t]$, starting with \cref{eq_app_lambda_for} and using the same expansion
\begin{align}
\label{eq_app_lam}
e^{t \Lambda_t}& = \ \langle e^{t h_t} \rangle_{\T{for}} \nn \\
& = \int e^{t h_t(s)} p_{\T{for}} (s,t) \di s \nn \\ 
& \underset{t \sigma \rightarrow 0}{\sim} e^{t \langle h_t \rangle_{\T{for}}} \int \lbk 1 + t  \lp h_t(s)-\langle h_t \rangle_{\T{for}} \rp +  \frac{t^2}{2} \lp h_t(s)-\langle h_t \rangle_{\T{for}} \rp^2 \rbk p_{\T{for}} (s,t) \di s \nn \\
& \sim e^{t \langle h_t \rangle_{\T{for}}} \lbk 1 + \frac{(t \sigma_{\T{for}}(h_t))^2}{2} \rbk\,,
\end{align}
which is a second-order correction to $\exp[t \langle h_t \rangle_{\T{for}}]$ in $t \sigma_{\T{for}}(h_t)$. 

Combining \cref{eq_app_h_back,eq_app_lam} we find at first order
\be
\label{eq_app_pi_lim}
\langle g_t \rangle_{\T{back}} - \langle g_t \rangle_{\T{for}} \underset{t \sigma \rightarrow 0}{\sim} t \ \mathrm{Cov_{\T{for}}} (h_t,g_t) \,.
\ee

As mentioned at the beginning of this section, we can use the backward point of view, with a first-order expansion in $t\sigma_{\T{back}}(h_t(s))$ instead. In this case,
\begin{align}
\label{eq_app_g_for}
\langle g_t \rangle_{\T{for}} &= e^{t (\Lambda_t - \langle h_t \rangle_{\T{back}})} \int g_t(s) e^{t (\langle h_t \rangle_{\T{back}} -h_t(s) )} p_{\T{back}} (s,t) \di s \nn \\
& \underset{t \sigma \rightarrow 0}{\sim} e^{t (\Lambda_t - \langle h_t \rangle_{\T{back}})} \lbk \langle g_t \rangle_{\T{back}} - t \mathrm{Cov_{\T{back}}} (h_t,g_t) \rbk \,.
\end{align}
The prefactor is computed with the same expansion starting from \cref{eq_app_lambda_back}:
\begin{align}
\label{eq_app_lam_back}
e^{-t \Lambda_t}& = \ \langle e^{-t h_t} \rangle_{\T{back}} \nn \\
& \underset{t \sigma \rightarrow 0}{\sim} e^{-t \langle h_t \rangle_{\T{back}}} \int \lbk 1 + t  \lp\langle h_t \rangle_{\T{back}} - h_t(s) \rp +  \frac{t^2}{2} \lp \langle h_t \rangle_{\T{back}} - h_t(s)\rp^2 \rbk p_{\T{back}} (s,t) \di s \nn \\
& \sim e^{-t \langle h_t \rangle_{\T{back}}} \lbk 1 + \frac{(t \sigma_{\T{back}}(h_t))^2}{2} \rbk\,,
\end{align}
which is a second order correction in $t \sigma_{\T{back}}(h_t)$. Combining \cref{eq_app_g_for,eq_app_lam_back}, we find 
\be
\label{eq_app_back_lim}
\langle g_t \rangle_{\T{back}} - \langle g_t \rangle_{\T{for}} \underset{t \sigma \rightarrow 0}{\sim} t \ \mathrm{Cov_{\T{back}}} (h_t,g_t) \,.
\ee
When comparing \cref{eq_app_pi_lim,eq_app_back_lim}, we conclude that the covariance can be taken equivalently in the forward or backward statistics.

Let us now turn to the r.h.s. of inequality \cref{eq_tur_s_min}. Using the expression of $\sigma_{\T{for}}(q_t)$ as the forward coefficient of variation of the quantity $\exp [t h_t(s)]$ (\cref{eq_std_q}) (resp. $\sigma_{\T{back}}(r_t)$ as the backward coefficient of variation of the quantity $\exp [-t h_t(s)]$), it is straight-forward to show from the same kind of Taylor expansion that 
\begin{align}
\sigma_{\T{for}}(g_t) \sigma_{\T{for}} (q_t) & \underset{t \sigma \rightarrow 0}{\sim} t \sigma_{\T{for}}(h_t) \sigma_{\T{for}}(g_t) \\
\sigma_{\T{back}}(g_t) \sigma_{\T{back}} (r_t) & \underset{t \sigma \rightarrow 0}{\sim} t \sigma_{\T{back}}(h_t) \sigma_{\T{back}}(g_t) \,.
\end{align}
Thus, the inequality \cref{eq_tur_s_min} does not get necessarily saturated in this limit. However,
in the particular case where $g_t(s)$ is the fitness landscape $h_t(s)$, then \cref{eq_app_pi_lim} reads
\be
\Pi_{\mathcal{S}} \underset{t \sigma \rightarrow 0}{\sim} t \ \mathrm{Var} (h_t) \,,
\ee
and thus the inequality \cref{eq_tur_hs} is saturated in this limit.

\section{A lower bound for the strength of selection}
\label{sec_app_low_bound}

\begin{figure*}
	\includegraphics[width=\linewidth]{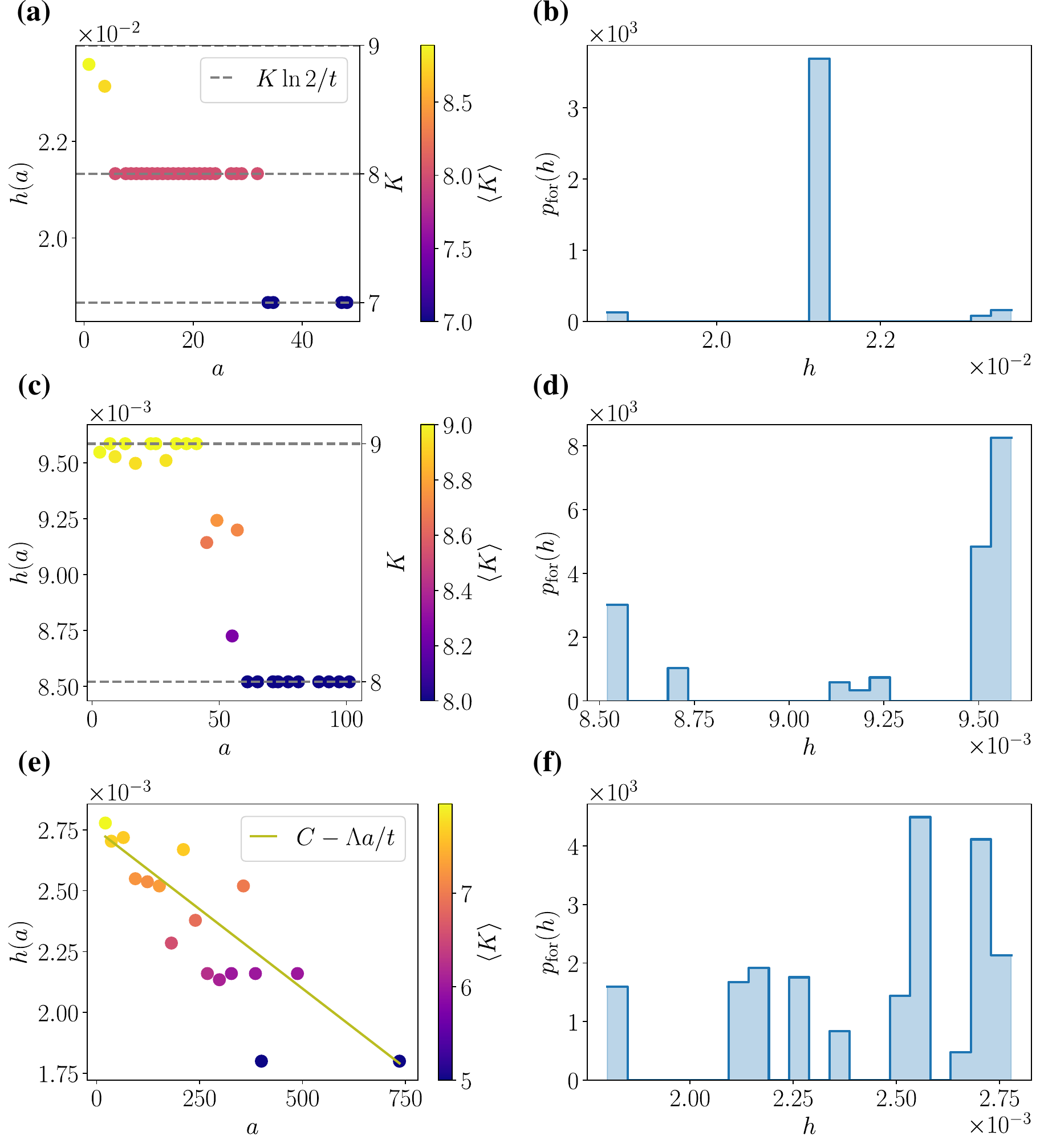}
	\caption{Experimental fitness landscapes for age and their forward distributions, computed with data from \cite{kiviet_stochasticity_2014}. Each line corresponds to a different experiment and the first column shows fitness landscapes $h_t(a)$ as functions of size $a$.
		\textbf{(a,c)}: the grey horizontal dashed lines correspond to theoretical plateaus, equal to $K \ln 2 /t$, predicted  when $K$ is fully determined by the value $s$ of the trait. The integers $K$ corresponding to the plateaus are indicated on the right y-axis.
		\textbf{(e)}: the plateaus are blurred and replaced by a smoother scatter plot in good agreement with the general shape of the theoretical prediction, made in the case of an age-controlled model in steady-state \cite{genthon_fluctuation_2020}. $\Lambda$ is the population growth rate and the constant $C$ was adjusted to fit the scatter plot.
		\textbf{(a,c,e)}: each dot is made of all the cells having the same age, and the mean number of divisions amongst those cells is represented by the color of the dot. This shows that dots aligning on a plateau corresponding to a number $K$ of divisions truly come from cells that underwent $K$ divisions.
		\textbf{(b,d,f)}: the second column represents the distribution $p_{\rm{for}}(h)$ of the corresponding age fitness landscapes (i.e. on the same line) with the forward age distribution. For all three rows, the fitness landscapes are highly non-Gaussian, and for \textbf{(b,d)} the peaks in these distributions correspond to the values of some of the plateaus.}	
	\label{fig_app_ha}
\end{figure*}

The trivial lower bound on the strength of selection $\Pi_{\mathcal{S}} \geq 0$ comes from 
the positivity of KL divergences $\mathcal{D}_{\rm{KL}}(p_{\T{for}} | p_{\T{back}})$ and $\mathcal{D}_{\rm{KL}}(p_{\T{back}} | p_{\T{for}})$, giving respectively $\Lambda_t - \langle h_t \rangle_\T{for}  \geq 0$ and $\langle h_t \rangle_{\T{back}} - \Lambda_t \geq 0$. 
The positivity of KL divergence itself relies on Jensen's inequality.
In order to improve upon this bound, we proceed in two steps, using a sharpened version of Jensen's inequality \cite{liao_sharpening_2019} to both inequalities.
First, seeking a positive lower bound for $\Lambda_t - \langle h_t \rangle_{\T{for}}$, the sharpened version of Jensen's inequality reads
\begin{equation}
\label{eq_app_jensen_for}
\langle e^{th_t} \rangle_{\T{for}} -  e^{t \langle h_t \rangle_{\T{for}}} \geq \sigma^2_{\T{for}}(h_t) \ \inf_{h} \psi(\varphi_{\T{for}},h,\langle h_t \rangle_{\T{for}}) \,,
\end{equation}
where functions $\psi$, $\varphi_{\T{for}}$ and $\varphi_{\T{back}}$ have been defined in the main text. The r.h.s involves the minimum of the function $\psi$ when varying $h$ on its support at time $t$. We then divide this expression by $\exp(t \Lambda_t)$ 
\begin{equation}
\langle e^{t(h_t-\Lambda_t)} \rangle_{\T{for}} -  e^{t (\langle h_t  \rangle_{\T{for}} - \Lambda_t)} \geq \frac{\sigma^2_{\T{for}}(h_t)}{\exp(t \Lambda_t)} \ \inf_{h} \psi(\varphi_{\T{for}},h,\langle h_t \rangle_{\T{for}}) \,,
\end{equation}
The first term is $1$ because of the normalization of the probability distribution $p_{\T{back}}$. Finally the enhanced bound reads
\begin{equation}
\label{eq-ineq1}
\Lambda_t - \langle h_t \rangle_{\T{for}} \geq 
- \frac{1}{t}\ln \lp 1 - \frac{\sigma^2_{\T{for}}(h_t)}{\exp(t \Lambda_t)} \ \inf_{h} \psi(\varphi_{\T{for}},h,\langle h_t \rangle_{\T{for}}) \rp \,.
\end{equation}
Similarly, we find
\begin{equation}
\label{eq-ineq2}
\langle h_t \rangle_{\T{back}} - \Lambda_t \geq \\ 
- \frac{1}{t} \ln \lp 1 - \frac{\sigma^2_{\T{back}}(h_t)}{\exp(-t \Lambda_t)} \ \inf_h \psi(\varphi_{\T{back}},h,\langle h_t \rangle_{\T{back}}) \rp \,.
\end{equation}
Liao et al. proved \cite{liao_sharpening_2019} that when $\varphi'(x)$ is a convex (resp. concave) function, then $\psi(\varphi,x,\nu)$ is an increasing (resp. decreasing) function of $x$, and thus the infimum of function $\psi(\varphi,x,\nu)$ on $x$ is reached for $x=x_{\T{min}}$ (resp. $x=x_{\T{max}}$).

Because of the convexity of $\varphi'_{\T{for}}(x)=t \exp \lbk tx \rbk$, the minimum of $\psi$ is reached when evaluating $\psi$ at the minimum value $h_{\T{min}}$ of the fitness landscape $h_t(s)$. At finite time, the support of $h_t(s)$ is finite and so is its minimum value. 
Similarly, because of the concavity of $\varphi'_{\T{back}}(x)=-t \exp \lbk -tx \rbk$, the minimum of $\psi$ is reached when evaluating $\psi$ at the maximum value $h_{\T{max}}$.
Finally, we use the relation $-\ln(1-x) \geq x$ valid for any real number $x$ and we combine the two inequalities to obtain 
\begin{equation}
\label{eq-lower bound}
\Pi_{\mathcal{S}} \ \geq \ \frac{1}{t} \Big[ \frac{\sigma^2_{\T{for}}(h_t)}{\exp(t \Lambda_t)} \ \psi(\varphi_{\T{for}}, h_{\T{min}},\langle h_t \rangle_{\T{for}})  
+ \frac{\sigma^2_{\T{back}}(h_t)}{\exp(-t \Lambda_t)} \ \psi(\varphi_{\T{back}}, h_{\T{max}},\langle h_t \rangle_{\T{back}}) \Big] \,.
\end{equation}
Note that the r.h.s. of \cref{eq-ineq1,eq-ineq2} are both positive numbers due to the convexity and concavity of 
$\varphi'_{\T{for}}$ and $\varphi'_{\T{back}}$ respectively. As a result, their sum is also positive and therefore the r.h.s. of \cref{eq-lower bound} does represent an improvement with respect to the trivial bound which would be $0$.

Liao et al. also proposed another lower bound, looser but simpler than the one involving the function $\psi$. Indeed, one can replace $\inf_{h} \psi(\varphi_{\T{for}},h,\langle h_t \rangle_{\T{for}})$ by $ \inf_{h} \varphi_{\T{for}}''(h)/2$ in \cref{eq_app_jensen_for}. Moreover $ \inf_{h} \varphi_{\T{for}}''(h)/2 = \varphi_{\T{for}}''(h_{\T{min}})/2$ since $\varphi_{\T{for}}''(h)$ is an increasing function of $h$. The same goes for the other inequality, and combining the two leads to

\begin{equation}
\label{eq-lower bound_v2}
\Pi_{\mathcal{S}} \ \geq \ \frac{1}{2t} \Big[ \frac{\sigma^2_{\T{for}}(h_t)}{\exp(t \Lambda_t)} \ \varphi_{\T{for}}''(h_{\T{min}})
+ \frac{\sigma^2_{\T{back}}(h_t)}{\exp(-t \Lambda_t)} \ \varphi_{\T{back}}''(h_{\T{max}}) \Big] \,.
\end{equation}
We notice that this version of the bound does not depend on the average values of the fitness landscape, unlike \cref{eq-lower bound}.

Let us mention that if no information is known on the support of the fitness landscape, $h_{\T{min}}$ can still be taken equal to $0$, in both \cref{eq-lower bound,eq-lower bound_v2}, because fitness landscapes are positive functions. Indeed, the fitness landscape can be expressed as \cite{nozoe_inferring_2017}:	$t h_t(s)=\ln \lbk \sum_K m^K R_{\T{for}}(K|s) \rbk$, 
in terms of the conditional forward probability $R_{\T{for}}(K|s)$ of the number of divisions. 
Since $m \geq 1$ and $K\geq 0$, this relation implies that the fitness landscape is a positive quantity.
In this case, \cref{eq-lower bound} (resp. \cref{eq-lower bound_v2}) gives a non-trivial bound based on the first two moments (resp. second moment) of the forward fitness landscape distribution.

The simplest bound is thus obtained when considering \cref{eq-lower bound_v2} with $h_{\T{min}}=0$ and $h_{\T{max}}=+\infty$, which cancels the second term in the bracket
\begin{equation}
\label{eq-lower bound_v3}
\Pi_{\mathcal{S}} \ \geq \ \frac{t}{2} \frac{\sigma^2_{\T{for}}(h_t)}{\exp(t \Lambda_t)} \,.
\end{equation}

\end{widetext}

\section{Further tests on experimental data from \cite{kiviet_stochasticity_2014}}
\label{sec_app_h_age}

\begin{figure}
\includegraphics[width=\linewidth]{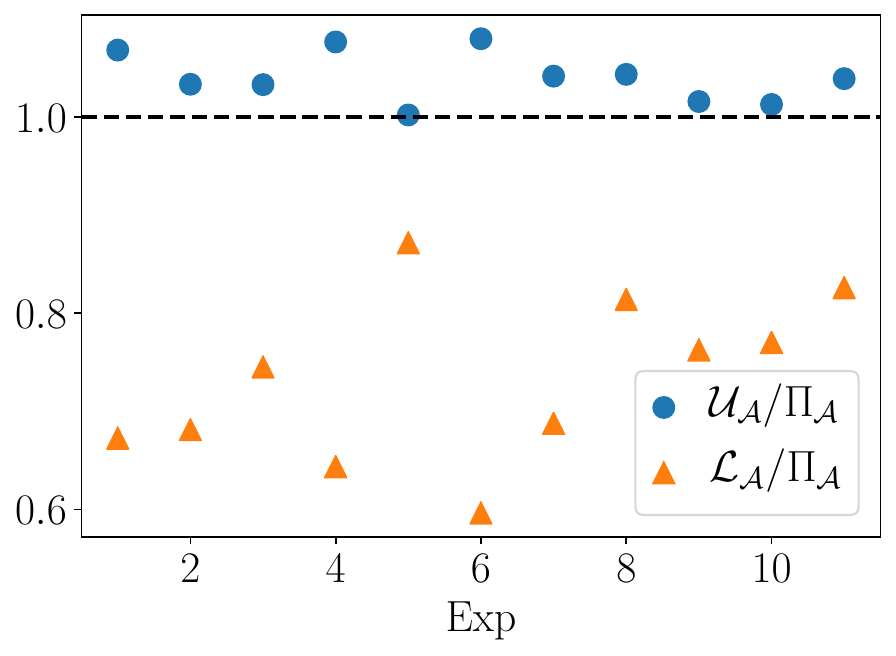}
\caption{Upper bound $\mathcal{U}_{\mathcal{A}}$ (blue dots) and lower bound $\mathcal{L}_{\mathcal{A}}$ (orange triangles) for the strength of selection acting on size $\Pi_{\mathcal{A}}$, normalized by the latter. The $x$-axis represents the $11$ colonies in different growth conditions from \cite{kiviet_stochasticity_2014}, in the same order as on \cref{fig_pix_bounds}. }
\label{fig_app_pia_bounds}
\end{figure}
On \cref{fig_hx} in \cref{sec_test}, we showed the size fitness landscapes $h_t(x)$ as a function of the cell size $x$, and the distribution $p_{\rm{for}}(h)$ of size fitness landscape computed with the forward cell size distribution, for three of the eleven experiments from \cite{kiviet_stochasticity_2014}.
We now show the corresponding plots when choosing the age $\mathcal{A}$ of the cell as the phenotypic trait $\mathcal{S}$. The experiment on line $i$ on \cref{fig_app_ha} is the same as the experiment on line $i$ on \cref{fig_hx}. 
By comparing the two figures, we see that for the first two rows the theoretical plateaus at $h=K \ln2 /t$ are the same for $h_t(a)$ and for $h_t(x)$, which is logical since it is the same cells, and both the age and the size are highly correlated to the number of divisions. 
On \cref{fig_app_ha}e, as for the size on \cref{fig_hx}e, the plateaus start to blur to give rise to a smoother scatter plot, whose shape matches the linear prediction we made for age-controlled models in steady state \cite{genthon_fluctuation_2020}. 
Similarly to the case of the cell size, distributions of fitness landscapes shown on the second column are highly non Gaussian. 
\begin{figure}
	\includegraphics[width=\linewidth]{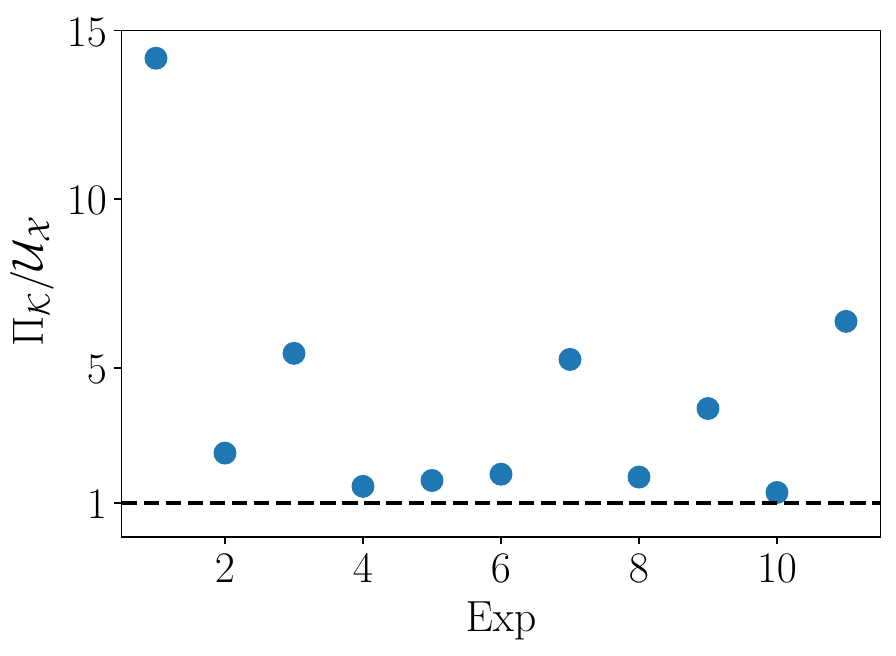}
	\includegraphics[width=\linewidth]{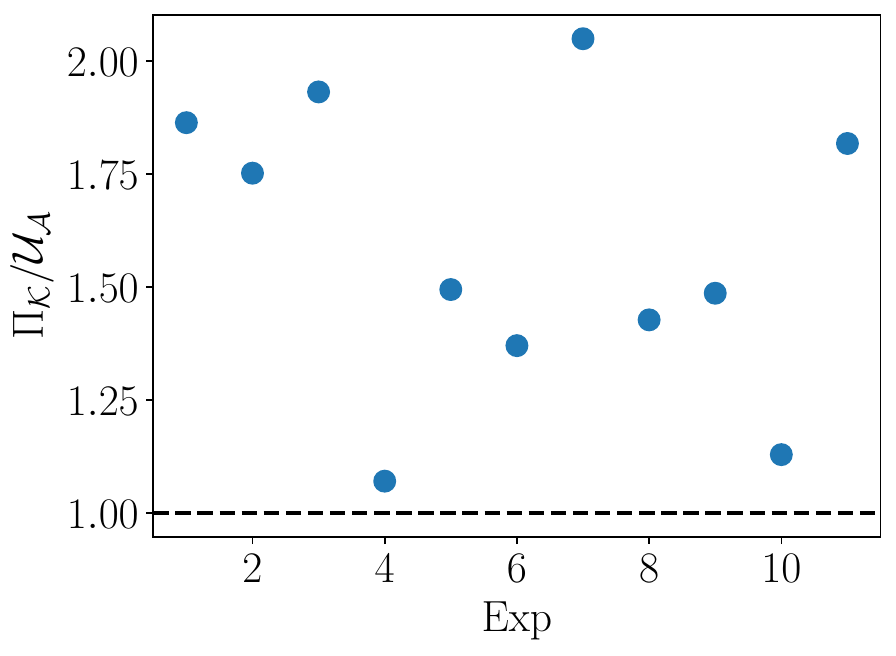}
	\caption{Ratio of the general bound $\Pi_{\mathcal{K}}$ to our upper bound $\mathcal{U}_{\mathcal{X}}$ for size (top plot) and $\mathcal{U}_{\mathcal{A}}$ for age (bottom plot) for the 11 experiments from \cite{kiviet_stochasticity_2014}, in no particular order. All points are above the black horizontal dashed line at $y=1$, which indicates that our upper bound is always smaller and thus better than $\Pi_{\mathcal{K}}$.}
	\label{fig_app_up_vs_pik}
\end{figure}

Then, we test the upper and lower bounds on the strength of selection acting on cell age using the same data. 
We show on \cref{fig_app_pia_bounds} the upper bound $\mathcal{U}_{\mathcal{A}}$ given by \cref{eq_tur_hs} and the lower bound $\mathcal{L}_{\mathcal{A}}$ given by \cref{eq_pi_low}, normalized by the strength of selection $\Pi_{\mathcal{A}}$ acting on age. The $x$-axis numbers the colonies which have grown in different nutrient medium \cite{kiviet_stochasticity_2014}. As expected, points representing the upper bound and those representing the lower bound are respectively above and below the horizontal dashed line at $y=1$. 	

Nozoe et al. proved that the strength of selection acting on any trait $\mathcal{S}$ is bounded by the strength of selection acting on the number of divisions $\mathcal{K}$ \cite{nozoe_inferring_2017}: $\forall \mathcal{S}, \Pi_{\mathcal{S}} \leq \Pi_{\mathcal{K}}$. We obtained in this paper a trait-dependent bound, which highlights the role of fluctuations of fitness landscape of that particular trait for the strength of selection, and which is often tighter than $\Pi_{\mathcal{K}}$. 
We show on \cref{fig_app_up_vs_pik} the ratio of $\Pi_{\mathcal{K}}$ to the upper bound $\mathcal{U}_{\mathcal{X}}$ for the size (top plot) and $\mathcal{U}_{\mathcal{A}}$ for the age (bottom plot). All points are indeed above $1$.


\begin{thebibliography}{34}%
	\makeatletter
	\providecommand \@ifxundefined [1]{%
		\@ifx{#1\undefined}
	}%
	\providecommand \@ifnum [1]{%
		\ifnum #1\expandafter \@firstoftwo
		\else \expandafter \@secondoftwo
		\fi
	}%
	\providecommand \@ifx [1]{%
		\ifx #1\expandafter \@firstoftwo
		\else \expandafter \@secondoftwo
		\fi
	}%
	\providecommand \natexlab [1]{#1}%
	\providecommand \enquote  [1]{``#1''}%
	\providecommand \bibnamefont  [1]{#1}%
	\providecommand \bibfnamefont [1]{#1}%
	\providecommand \citenamefont [1]{#1}%
	\providecommand \href@noop [0]{\@secondoftwo}%
	\providecommand \href [0]{\begingroup \@sanitize@url \@href}%
	\providecommand \@href[1]{\@@startlink{#1}\@@href}%
	\providecommand \@@href[1]{\endgroup#1\@@endlink}%
	\providecommand \@sanitize@url [0]{\catcode `\\12\catcode `\$12\catcode
		`\&12\catcode `\#12\catcode `\^12\catcode `\_12\catcode `\%12\relax}%
	\providecommand \@@startlink[1]{}%
	\providecommand \@@endlink[0]{}%
	\providecommand \url  [0]{\begingroup\@sanitize@url \@url }%
	\providecommand \@url [1]{\endgroup\@href {#1}{\urlprefix }}%
	\providecommand \urlprefix  [0]{URL }%
	\providecommand \Eprint [0]{\href }%
	\providecommand \doibase [0]{https://doi.org/}%
	\providecommand \selectlanguage [0]{\@gobble}%
	\providecommand \bibinfo  [0]{\@secondoftwo}%
	\providecommand \bibfield  [0]{\@secondoftwo}%
	\providecommand \translation [1]{[#1]}%
	\providecommand \BibitemOpen [0]{}%
	\providecommand \bibitemStop [0]{}%
	\providecommand \bibitemNoStop [0]{.\EOS\space}%
	\providecommand \EOS [0]{\spacefactor3000\relax}%
	\providecommand \BibitemShut  [1]{\csname bibitem#1\endcsname}%
	\let\auto@bib@innerbib\@empty
	\bibitem [{\citenamefont {Leibler}\ and\ \citenamefont
		{Kussell}(2010)}]{leibler_individual_2010}%
	\BibitemOpen
	\bibfield  {author} {\bibinfo {author} {\bibfnamefont {S.}~\bibnamefont
			{Leibler}}\ and\ \bibinfo {author} {\bibfnamefont {E.}~\bibnamefont
			{Kussell}},\ }\bibfield  {title} {\bibinfo {title} {Individual histories and
			selection in heterogeneous populations},\ }\href
	{https://doi.org/10.1073/pnas.0912538107} {\bibfield  {journal} {\bibinfo
			{journal} {Proc. Natl. Acad. Sci. U.S.A.}\ }\textbf {\bibinfo {volume}
			{107}},\ \bibinfo {pages} {13183} (\bibinfo {year} {2010})}\BibitemShut
	{NoStop}%
	\bibitem [{\citenamefont {Nozoe}\ \emph {et~al.}(2017)\citenamefont {Nozoe},
		\citenamefont {Kussell},\ and\ \citenamefont
		{Wakamoto}}]{nozoe_inferring_2017}%
	\BibitemOpen
	\bibfield  {author} {\bibinfo {author} {\bibfnamefont {T.}~\bibnamefont
			{Nozoe}}, \bibinfo {author} {\bibfnamefont {E.}~\bibnamefont {Kussell}},\
		and\ \bibinfo {author} {\bibfnamefont {Y.}~\bibnamefont {Wakamoto}},\
	}\bibfield  {title} {\bibinfo {title} {Inferring fitness landscapes and
			selection on phenotypic states from single-cell genealogical data},\ }\href
	{https://doi.org/10.1371/journal.pgen.1006653} {\bibfield  {journal}
		{\bibinfo  {journal} {PLoS Genet.}\ }\textbf {\bibinfo {volume} {13}},\
		\bibinfo {pages} {e1006653} (\bibinfo {year} {2017})}\BibitemShut {NoStop}%
	\bibitem [{\citenamefont {Wakamoto}\ \emph {et~al.}(2012)\citenamefont
		{Wakamoto}, \citenamefont {Grosberg},\ and\ \citenamefont
		{Kussell}}]{wakamoto_optimal_2012}%
	\BibitemOpen
	\bibfield  {author} {\bibinfo {author} {\bibfnamefont {Y.}~\bibnamefont
			{Wakamoto}}, \bibinfo {author} {\bibfnamefont {A.~Y.}\ \bibnamefont
			{Grosberg}},\ and\ \bibinfo {author} {\bibfnamefont {E.}~\bibnamefont
			{Kussell}},\ }\bibfield  {title} {\bibinfo {title} {Optimal lineage principle
			for age-structured populations},\ }\href
	{https://doi.org/10.1111/j.1558-5646.2011.01418.x} {\bibfield  {journal}
		{\bibinfo  {journal} {Evolution}\ }\textbf {\bibinfo {volume} {66}},\
		\bibinfo {pages} {115} (\bibinfo {year} {2012})}\BibitemShut {NoStop}%
	\bibitem [{\citenamefont {Lambert}\ and\ \citenamefont
		{Kussell}(2015)}]{lambert_quantifying_2015}%
	\BibitemOpen
	\bibfield  {author} {\bibinfo {author} {\bibfnamefont {G.}~\bibnamefont
			{Lambert}}\ and\ \bibinfo {author} {\bibfnamefont {E.}~\bibnamefont
			{Kussell}},\ }\bibfield  {title} {\bibinfo {title} {Quantifying {Selective}
			{Pressures} {Driving} {Bacterial} {Evolution} {Using} {Lineage} {Analysis}},\
	}\href {https://doi.org/10.1103/PhysRevX.5.011016} {\bibfield  {journal}
		{\bibinfo  {journal} {Phys. Rev. X}\ }\textbf {\bibinfo {volume} {5}},\
		\bibinfo {pages} {011016} (\bibinfo {year} {2015})}\BibitemShut {NoStop}%
	\bibitem [{\citenamefont {Lin}\ and\ \citenamefont
		{Amir}(2017)}]{lin_effects_2017}%
	\BibitemOpen
	\bibfield  {author} {\bibinfo {author} {\bibfnamefont {J.}~\bibnamefont
			{Lin}}\ and\ \bibinfo {author} {\bibfnamefont {A.}~\bibnamefont {Amir}},\
	}\bibfield  {title} {\bibinfo {title} {The {Effects} of {Stochasticity} at
			the {Single}-{Cell} {Level} and {Cell} {Size} {Control} on the {Population}
			{Growth}},\ }\href {https://doi.org/10.1016/j.cels.2017.08.015} {\bibfield
		{journal} {\bibinfo  {journal} {Cell Syst.}\ }\textbf {\bibinfo {volume}
			{5}},\ \bibinfo {pages} {358} (\bibinfo {year} {2017})}\BibitemShut {NoStop}%
	\bibitem [{\citenamefont {Mustonen}\ and\ \citenamefont
		{Lassig}(2010)}]{mustonen_fitness_2010}%
	\BibitemOpen
	\bibfield  {author} {\bibinfo {author} {\bibfnamefont {V.}~\bibnamefont
			{Mustonen}}\ and\ \bibinfo {author} {\bibfnamefont {M.}~\bibnamefont
			{Lassig}},\ }\bibfield  {title} {\bibinfo {title} {Fitness flux and ubiquity
			of adaptive evolution},\ }\href {https://doi.org/10.1073/pnas.0907953107}
	{\bibfield  {journal} {\bibinfo  {journal} {Proc. Natl. Acad. Sci. U.S.A.}\
		}\textbf {\bibinfo {volume} {107}},\ \bibinfo {pages} {4248} (\bibinfo {year}
		{2010})}\BibitemShut {NoStop}%
	\bibitem [{\citenamefont {Mustonen}\ and\ \citenamefont
		{Lässig}(2009)}]{mustonen_fitness_2009}%
	\BibitemOpen
	\bibfield  {author} {\bibinfo {author} {\bibfnamefont {V.}~\bibnamefont
			{Mustonen}}\ and\ \bibinfo {author} {\bibfnamefont {M.}~\bibnamefont
			{Lässig}},\ }\bibfield  {title} {\bibinfo {title} {From fitness landscapes
			to seascapes: non-equilibrium dynamics of selection and adaptation},\ }\href
	{https://doi.org/10.1016/j.tig.2009.01.002} {\bibfield  {journal} {\bibinfo
			{journal} {Trends Genet.}\ }\textbf {\bibinfo {volume} {25}},\ \bibinfo
		{pages} {111} (\bibinfo {year} {2009})}\BibitemShut {NoStop}%
	\bibitem [{\citenamefont {Kobayashi}\ and\ \citenamefont
		{Sughiyama}(2015)}]{kobayashi_fluctuation_2015}%
	\BibitemOpen
	\bibfield  {author} {\bibinfo {author} {\bibfnamefont {T.~J.}\ \bibnamefont
			{Kobayashi}}\ and\ \bibinfo {author} {\bibfnamefont {Y.}~\bibnamefont
			{Sughiyama}},\ }\bibfield  {title} {\bibinfo {title} {Fluctuation {Relations}
			of {Fitness} and {Information} in {Population} {Dynamics}},\ }\href
	{https://doi.org/10.1103/PhysRevLett.115.238102} {\bibfield  {journal}
		{\bibinfo  {journal} {Phys. Rev. Lett.}\ }\textbf {\bibinfo {volume} {115}},\
		\bibinfo {pages} {238102} (\bibinfo {year} {2015})}\BibitemShut {NoStop}%
	\bibitem [{\citenamefont {García-García}\ \emph {et~al.}(2019)\citenamefont
		{García-García}, \citenamefont {Genthon},\ and\ \citenamefont
		{Lacoste}}]{garcia-garcia_linking_2019}%
	\BibitemOpen
	\bibfield  {author} {\bibinfo {author} {\bibfnamefont {R.}~\bibnamefont
			{García-García}}, \bibinfo {author} {\bibfnamefont {A.}~\bibnamefont
			{Genthon}},\ and\ \bibinfo {author} {\bibfnamefont {D.}~\bibnamefont
			{Lacoste}},\ }\bibfield  {title} {\bibinfo {title} {Linking lineage and
			population observables in biological branching processes},\ }\href
	{https://doi.org/10.1103/PhysRevE.99.042413} {\bibfield  {journal} {\bibinfo
			{journal} {Phys. Rev. E}\ }\textbf {\bibinfo {volume} {99}},\ \bibinfo
		{pages} {042413} (\bibinfo {year} {2019})}\BibitemShut {NoStop}%
	\bibitem [{\citenamefont {Genthon}\ and\ \citenamefont
		{Lacoste}(2020)}]{genthon_fluctuation_2020}%
	\BibitemOpen
	\bibfield  {author} {\bibinfo {author} {\bibfnamefont {A.}~\bibnamefont
			{Genthon}}\ and\ \bibinfo {author} {\bibfnamefont {D.}~\bibnamefont
			{Lacoste}},\ }\bibfield  {title} {\bibinfo {title} {Fluctuation relations and
			fitness landscapes of growing cell populations},\ }\href
	{https://doi.org/10.1038/s41598-020-68444-x} {\bibfield  {journal} {\bibinfo
			{journal} {Sci. Rep.}\ }\textbf {\bibinfo {volume} {10}},\ \bibinfo {pages}
		{11889} (\bibinfo {year} {2020})}\BibitemShut {NoStop}%
	\bibitem [{\citenamefont {Boyer}\ \emph {et~al.}(2016)\citenamefont {Boyer},
		\citenamefont {Biswas}, \citenamefont {Kumar~Soshee}, \citenamefont
		{Scaramozzino}, \citenamefont {Nizak},\ and\ \citenamefont
		{Rivoire}}]{boyer_hierarchy_2016}%
	\BibitemOpen
	\bibfield  {author} {\bibinfo {author} {\bibfnamefont {S.}~\bibnamefont
			{Boyer}}, \bibinfo {author} {\bibfnamefont {D.}~\bibnamefont {Biswas}},
		\bibinfo {author} {\bibfnamefont {A.}~\bibnamefont {Kumar~Soshee}}, \bibinfo
		{author} {\bibfnamefont {N.}~\bibnamefont {Scaramozzino}}, \bibinfo {author}
		{\bibfnamefont {C.}~\bibnamefont {Nizak}},\ and\ \bibinfo {author}
		{\bibfnamefont {O.}~\bibnamefont {Rivoire}},\ }\bibfield  {title} {\bibinfo
		{title} {Hierarchy and extremes in selections from pools of randomized
			proteins},\ }\href {https://doi.org/10.1073/pnas.1517813113} {\bibfield
		{journal} {\bibinfo  {journal} {Proc. Natl. Acad. Sci. U.S.A.}\ }\textbf
		{\bibinfo {volume} {113}},\ \bibinfo {pages} {3482} (\bibinfo {year}
		{2016})}\BibitemShut {NoStop}%
	\bibitem [{\citenamefont {Smerlak}\ and\ \citenamefont
		{Youssef}(2017)}]{smerlak_limiting_2017}%
	\BibitemOpen
	\bibfield  {author} {\bibinfo {author} {\bibfnamefont {M.}~\bibnamefont
			{Smerlak}}\ and\ \bibinfo {author} {\bibfnamefont {A.}~\bibnamefont
			{Youssef}},\ }\bibfield  {title} {\bibinfo {title} {Limiting fitness
			distributions in evolutionary dynamics},\ }\href
	{https://doi.org/10.1016/j.jtbi.2017.01.005} {\bibfield  {journal} {\bibinfo
			{journal} {J. Theor. Biol.}\ }\textbf {\bibinfo {volume} {416}},\ \bibinfo
		{pages} {68} (\bibinfo {year} {2017})}\BibitemShut {NoStop}%
	\bibitem [{\citenamefont {Dechant}\ and\ \citenamefont
		{Sasa}(2020)}]{dechant_fluctuationresponse_2020}%
	\BibitemOpen
	\bibfield  {author} {\bibinfo {author} {\bibfnamefont {A.}~\bibnamefont
			{Dechant}}\ and\ \bibinfo {author} {\bibfnamefont {S.-i.}\ \bibnamefont
			{Sasa}},\ }\bibfield  {title} {\bibinfo {title} {Fluctuation–response
			inequality out of equilibrium},\ }\href
	{https://doi.org/10.1073/pnas.1918386117} {\bibfield  {journal} {\bibinfo
			{journal} {Proc. Natl. Acad. Sci. U.S.A.}\ }\textbf {\bibinfo {volume}
			{117}},\ \bibinfo {pages} {6430} (\bibinfo {year} {2020})}\BibitemShut
	{NoStop}%
	\bibitem [{\citenamefont {Dinis}\ \emph {et~al.}(2020)\citenamefont {Dinis},
		\citenamefont {Unterberger},\ and\ \citenamefont
		{Lacoste}}]{dinis_phase_2020}%
	\BibitemOpen
	\bibfield  {author} {\bibinfo {author} {\bibfnamefont {L.}~\bibnamefont
			{Dinis}}, \bibinfo {author} {\bibfnamefont {J.}~\bibnamefont {Unterberger}},\
		and\ \bibinfo {author} {\bibfnamefont {D.}~\bibnamefont {Lacoste}},\
	}\bibfield  {title} {\bibinfo {title} {Phase transitions in optimal betting
			strategies},\ }\href {https://doi.org/10.1209/0295-5075/131/60005} {\bibfield
		{journal} {\bibinfo  {journal} {EPL}\ }\textbf {\bibinfo {volume} {131}},\
		\bibinfo {pages} {60005} (\bibinfo {year} {2020})}\BibitemShut {NoStop}%
	\bibitem [{\citenamefont {Horowitz}\ and\ \citenamefont
		{Gingrich}(2020)}]{horowitz_thermodynamic_2020}%
	\BibitemOpen
	\bibfield  {author} {\bibinfo {author} {\bibfnamefont {J.~M.}\ \bibnamefont
			{Horowitz}}\ and\ \bibinfo {author} {\bibfnamefont {T.~R.}\ \bibnamefont
			{Gingrich}},\ }\bibfield  {title} {\bibinfo {title} {Thermodynamic
			uncertainty relations constrain non-equilibrium fluctuations},\ }\href
	{https://doi.org/10.1038/s41567-019-0702-6} {\bibfield  {journal} {\bibinfo
			{journal} {Nat. Phys.}\ }\textbf {\bibinfo {volume} {16}},\ \bibinfo {pages}
		{15} (\bibinfo {year} {2020})}\BibitemShut {NoStop}%
	\bibitem [{\citenamefont {Thomas}(2017)}]{thomas_making_2017}%
	\BibitemOpen
	\bibfield  {author} {\bibinfo {author} {\bibfnamefont {P.}~\bibnamefont
			{Thomas}},\ }\bibfield  {title} {\bibinfo {title} {Making sense of snapshot
			data: ergodic principle for clonal cell populations},\ }\href
	{https://doi.org/10.1098/rsif.2017.0467} {\bibfield  {journal} {\bibinfo
			{journal} {J. R. Soc. Interface}\ }\textbf {\bibinfo {volume} {14}},\
		\bibinfo {pages} {20170467} (\bibinfo {year} {2017})}\BibitemShut {NoStop}%
	\bibitem [{\citenamefont {Baake}\ and\ \citenamefont
		{Georgii}(2007)}]{baake_mutation_2007}%
	\BibitemOpen
	\bibfield  {author} {\bibinfo {author} {\bibfnamefont {E.}~\bibnamefont
			{Baake}}\ and\ \bibinfo {author} {\bibfnamefont {H.-O.}\ \bibnamefont
			{Georgii}},\ }\bibfield  {title} {\bibinfo {title} {Mutation, selection, and
			ancestry in branching models: a variational approach},\ }\href
	{https://doi.org/10.1007/s00285-006-0039-5} {\bibfield  {journal} {\bibinfo
			{journal} {J. Math. Biol.}\ }\textbf {\bibinfo {volume} {54}},\ \bibinfo
		{pages} {257} (\bibinfo {year} {2007})}\BibitemShut {NoStop}%
	\bibitem [{\citenamefont {Smerlak}(2021)}]{smerlak_quasi-species_2021}%
	\BibitemOpen
	\bibfield  {author} {\bibinfo {author} {\bibfnamefont {M.}~\bibnamefont
			{Smerlak}},\ }\bibfield  {title} {\bibinfo {title} {Quasi-species evolution
			maximizes genotypic reproductive value (not fitness or flatness)}\ }\href
	{https://doi.org/10.1101/2021.02.23.432496} {10.1101/2021.02.23.432496}
	(\bibinfo {year} {2021})\BibitemShut {NoStop}%
	\bibitem [{\citenamefont {Wang}\ \emph {et~al.}(2010)\citenamefont {Wang},
		\citenamefont {Robert}, \citenamefont {Pelletier}, \citenamefont {Dang},
		\citenamefont {Taddei}, \citenamefont {Wright},\ and\ \citenamefont
		{Jun}}]{wang_robust_2010}%
	\BibitemOpen
	\bibfield  {author} {\bibinfo {author} {\bibfnamefont {P.}~\bibnamefont
			{Wang}}, \bibinfo {author} {\bibfnamefont {L.}~\bibnamefont {Robert}},
		\bibinfo {author} {\bibfnamefont {J.}~\bibnamefont {Pelletier}}, \bibinfo
		{author} {\bibfnamefont {W.~L.}\ \bibnamefont {Dang}}, \bibinfo {author}
		{\bibfnamefont {F.}~\bibnamefont {Taddei}}, \bibinfo {author} {\bibfnamefont
			{A.}~\bibnamefont {Wright}},\ and\ \bibinfo {author} {\bibfnamefont
			{S.}~\bibnamefont {Jun}},\ }\bibfield  {title} {\bibinfo {title} {Robust
			{Growth} of {Escherichia} coli},\ }\href
	{https://doi.org/10.1016/j.cub.2010.04.045} {\bibfield  {journal} {\bibinfo
			{journal} {Curr. Biol.}\ }\textbf {\bibinfo {volume} {20}},\ \bibinfo {pages}
		{1099} (\bibinfo {year} {2010})}\BibitemShut {NoStop}%
	\bibitem [{\citenamefont {Wright}(1932)}]{wright_roles_1932}%
	\BibitemOpen
	\bibfield  {author} {\bibinfo {author} {\bibfnamefont {S.}~\bibnamefont
			{Wright}},\ }\bibfield  {title} {\bibinfo {title} {The roles of mutation,
			inbreeding, crossbreeding and selection in evolution},\ }\href@noop {}
	{\bibfield  {journal} {\bibinfo  {journal} {Proceedings of the Sixth
				International Congress on Genetics}\ }\textbf {\bibinfo {volume} {1}},\
		\bibinfo {pages} {356} (\bibinfo {year} {1932})}\BibitemShut {NoStop}%
	\bibitem [{\citenamefont {Peliti}(1997)}]{peliti_introduction_1997}%
	\BibitemOpen
	\bibfield  {author} {\bibinfo {author} {\bibfnamefont {L.}~\bibnamefont
			{Peliti}},\ }\bibfield  {title} {\bibinfo {title} {Introduction to the
			statistical theory of {Darwinian} evolution},\ }\href
	{http://arxiv.org/abs/cond-mat/9712027} {\bibfield  {journal} {\bibinfo
			{journal} {arXiv:cond-mat/9712027}\ } (\bibinfo {year} {1997})}\BibitemShut
	{NoStop}%
	\bibitem [{\citenamefont {Neher}\ and\ \citenamefont
		{Shraiman}(2011)}]{neher_statistical_2011}%
	\BibitemOpen
	\bibfield  {author} {\bibinfo {author} {\bibfnamefont {R.~A.}\ \bibnamefont
			{Neher}}\ and\ \bibinfo {author} {\bibfnamefont {B.~I.}\ \bibnamefont
			{Shraiman}},\ }\bibfield  {title} {\bibinfo {title} {Statistical genetics and
			evolution of quantitative traits},\ }\href
	{https://doi.org/10.1103/RevModPhys.83.1283} {\bibfield  {journal} {\bibinfo
			{journal} {Rev. Mod. Phys.}\ }\textbf {\bibinfo {volume} {83}},\ \bibinfo
		{pages} {1283} (\bibinfo {year} {2011})}\BibitemShut {NoStop}%
	\bibitem [{\citenamefont {Fisher}(2000)}]{fisher_genetical_2000}%
	\BibitemOpen
	\bibfield  {author} {\bibinfo {author} {\bibfnamefont {R.~A.}\ \bibnamefont
			{Fisher}},\ }\href@noop {} {\emph {\bibinfo {title} {The {Genetical} {Theory}
				of {Natural} {Selection}: {A} {Complete} {Variorum} {Edition}}}},\ \bibinfo
	{edition} {{Bennet} {J}. {H}.}\ ed.\ (\bibinfo  {publisher} {Oxford
		University Press},\ \bibinfo {address} {Oxford, UK},\ \bibinfo {year}
	{2000})\BibitemShut {NoStop}%
	\bibitem [{\citenamefont {Price}(1972)}]{price_fishers_1972}%
	\BibitemOpen
	\bibfield  {author} {\bibinfo {author} {\bibfnamefont {G.~R.}\ \bibnamefont
			{Price}},\ }\bibfield  {title} {\bibinfo {title} {Fisher's 'fundamental
			theorem' made clear},\ }\href
	{https://doi.org/10.1111/j.1469-1809.1972.tb00764.x} {\bibfield  {journal}
		{\bibinfo  {journal} {Ann. Hum. Genet.}\ }\textbf {\bibinfo {volume} {36}},\
		\bibinfo {pages} {129} (\bibinfo {year} {1972})}\BibitemShut {NoStop}%
	\bibitem [{\citenamefont {Gardner}(2020)}]{gardner_prices_2020}%
	\BibitemOpen
	\bibfield  {author} {\bibinfo {author} {\bibfnamefont {A.}~\bibnamefont
			{Gardner}},\ }\bibfield  {title} {\bibinfo {title} {Price's equation made
			clear},\ }\href {https://doi.org/10.1098/rstb.2019.0361} {\bibfield
		{journal} {\bibinfo  {journal} {Philos. Trans. R. Soc. B}\ }\textbf {\bibinfo
			{volume} {375}},\ \bibinfo {pages} {20190361} (\bibinfo {year}
		{2020})}\BibitemShut {NoStop}%
	\bibitem [{\citenamefont {Liao}\ and\ \citenamefont
		{Berg}(2019)}]{liao_sharpening_2019}%
	\BibitemOpen
	\bibfield  {author} {\bibinfo {author} {\bibfnamefont {J.~G.}\ \bibnamefont
			{Liao}}\ and\ \bibinfo {author} {\bibfnamefont {A.}~\bibnamefont {Berg}},\
	}\bibfield  {title} {\bibinfo {title} {Sharpening {Jensen}'s {Inequality}},\
	}\href {https://doi.org/10.1080/00031305.2017.1419145} {\bibfield  {journal}
		{\bibinfo  {journal} {Am. Stat.}\ }\textbf {\bibinfo {volume} {73}},\
		\bibinfo {pages} {278} (\bibinfo {year} {2019})}\BibitemShut {NoStop}%
	\bibitem [{\citenamefont {Kiviet}\ \emph {et~al.}(2014)\citenamefont {Kiviet},
		\citenamefont {Nghe}, \citenamefont {Walker}, \citenamefont {Boulineau},
		\citenamefont {Sunderlikova},\ and\ \citenamefont
		{Tans}}]{kiviet_stochasticity_2014}%
	\BibitemOpen
	\bibfield  {author} {\bibinfo {author} {\bibfnamefont {D.~J.}\ \bibnamefont
			{Kiviet}}, \bibinfo {author} {\bibfnamefont {P.}~\bibnamefont {Nghe}},
		\bibinfo {author} {\bibfnamefont {N.}~\bibnamefont {Walker}}, \bibinfo
		{author} {\bibfnamefont {S.}~\bibnamefont {Boulineau}}, \bibinfo {author}
		{\bibfnamefont {V.}~\bibnamefont {Sunderlikova}},\ and\ \bibinfo {author}
		{\bibfnamefont {S.~J.}\ \bibnamefont {Tans}},\ }\bibfield  {title} {\bibinfo
		{title} {Stochasticity of metabolism and growth at the single-cell level},\
	}\href {https://doi.org/10.1038/nature13582} {\bibfield  {journal} {\bibinfo
			{journal} {Nature}\ }\textbf {\bibinfo {volume} {514}},\ \bibinfo {pages}
		{376} (\bibinfo {year} {2014})}\BibitemShut {NoStop}%
	\bibitem [{\citenamefont {Kaneko}\ and\ \citenamefont
		{Furusawa}(2018)}]{kaneko_macroscopic_2018}%
	\BibitemOpen
	\bibfield  {author} {\bibinfo {author} {\bibfnamefont {K.}~\bibnamefont
			{Kaneko}}\ and\ \bibinfo {author} {\bibfnamefont {C.}~\bibnamefont
			{Furusawa}},\ }\bibfield  {title} {\bibinfo {title} {Macroscopic {Theory} for
			{Evolving} {Biological} {Systems} {Akin} to {Thermodynamics}},\ }\href
	{https://doi.org/10.1146/annurev-biophys-070317-033155} {\bibfield  {journal}
		{\bibinfo  {journal} {Annu. Rev. Biophys.}\ }\textbf {\bibinfo {volume}
			{47}},\ \bibinfo {pages} {273} (\bibinfo {year} {2018})}\BibitemShut
	{NoStop}%
	\bibitem [{\citenamefont {Schuh}\ and\ \citenamefont
		{Brower}(2009)}]{schuh_biological_2009}%
	\BibitemOpen
	\bibfield  {author} {\bibinfo {author} {\bibfnamefont {R.~T.}\ \bibnamefont
			{Schuh}}\ and\ \bibinfo {author} {\bibfnamefont {A.~V.~Z.}\ \bibnamefont
			{Brower}},\ }\href@noop {} {\emph {\bibinfo {title} {Biological systematics:
				principles and applications}}},\ \bibinfo {edition} {2nd}\ ed.\ (\bibinfo
	{publisher} {Comstock Pub. Associates/Cornell University Press},\ \bibinfo
	{address} {Ithaca},\ \bibinfo {year} {2009})\BibitemShut {NoStop}%
	\bibitem [{\citenamefont {Nguyen~Ba}\ \emph {et~al.}(2019)\citenamefont
		{Nguyen~Ba}, \citenamefont {Cvijović}, \citenamefont {Rojas~Echenique},
		\citenamefont {Lawrence}, \citenamefont {Rego-Costa}, \citenamefont {Liu},
		\citenamefont {Levy},\ and\ \citenamefont
		{Desai}}]{nguyen_ba_high-resolution_2019}%
	\BibitemOpen
	\bibfield  {author} {\bibinfo {author} {\bibfnamefont {A.~N.}\ \bibnamefont
			{Nguyen~Ba}}, \bibinfo {author} {\bibfnamefont {I.}~\bibnamefont
			{Cvijović}}, \bibinfo {author} {\bibfnamefont {J.~I.}\ \bibnamefont
			{Rojas~Echenique}}, \bibinfo {author} {\bibfnamefont {K.~R.}\ \bibnamefont
			{Lawrence}}, \bibinfo {author} {\bibfnamefont {A.}~\bibnamefont
			{Rego-Costa}}, \bibinfo {author} {\bibfnamefont {X.}~\bibnamefont {Liu}},
		\bibinfo {author} {\bibfnamefont {S.~F.}\ \bibnamefont {Levy}},\ and\
		\bibinfo {author} {\bibfnamefont {M.~M.}\ \bibnamefont {Desai}},\ }\bibfield
	{title} {\bibinfo {title} {High-resolution lineage tracking reveals
			travelling wave of adaptation in laboratory yeast},\ }\href
	{https://doi.org/10.1038/s41586-019-1749-3} {\bibfield  {journal} {\bibinfo
			{journal} {Nature}\ }\textbf {\bibinfo {volume} {575}},\ \bibinfo {pages}
		{494} (\bibinfo {year} {2019})}\BibitemShut {NoStop}%
	\bibitem [{\citenamefont {Tak}\ \emph {et~al.}(2019)\citenamefont {Tak},
		\citenamefont {Prevedello}, \citenamefont {Simon}, \citenamefont {Paillon},
		\citenamefont {Duffy},\ and\ \citenamefont {Perié}}]{tak_simultaneous_2019}%
	\BibitemOpen
	\bibfield  {author} {\bibinfo {author} {\bibfnamefont {T.}~\bibnamefont
			{Tak}}, \bibinfo {author} {\bibfnamefont {G.}~\bibnamefont {Prevedello}},
		\bibinfo {author} {\bibfnamefont {G.}~\bibnamefont {Simon}}, \bibinfo
		{author} {\bibfnamefont {N.}~\bibnamefont {Paillon}}, \bibinfo {author}
		{\bibfnamefont {K.~R.}\ \bibnamefont {Duffy}},\ and\ \bibinfo {author}
		{\bibfnamefont {L.}~\bibnamefont {Perié}},\ }\bibfield  {title} {\bibinfo
		{title} {Simultaneous tracking of division and differentiation from
			individual hematopoietic stem and progenitor cells reveals within-family
			homogeneity despite population heterogeneity}\ }\href
	{https://doi.org/10.1101/586354} {10.1101/586354} (\bibinfo {year}
	{2019})\BibitemShut {NoStop}%
	\bibitem [{\citenamefont {Goldenfeld}\ and\ \citenamefont
		{Woese}(2011)}]{goldenfeld_life_2011}%
	\BibitemOpen
	\bibfield  {author} {\bibinfo {author} {\bibfnamefont {N.}~\bibnamefont
			{Goldenfeld}}\ and\ \bibinfo {author} {\bibfnamefont {C.}~\bibnamefont
			{Woese}},\ }\bibfield  {title} {\bibinfo {title} {Life is {Physics}:
			{Evolution} as a {Collective} {Phenomenon} {Far} {From} {Equilibrium}},\
	}\href {https://doi.org/10.1146/annurev-conmatphys-062910-140509} {\bibfield
		{journal} {\bibinfo  {journal} {Annu. Rev. Condens. Matter Phys.}\ }\textbf
		{\bibinfo {volume} {2}},\ \bibinfo {pages} {375} (\bibinfo {year}
		{2011})}\BibitemShut {NoStop}%
	\bibitem [{\citenamefont {Powell}(1956)}]{powell_growth_1956}%
	\BibitemOpen
	\bibfield  {author} {\bibinfo {author} {\bibfnamefont {E.~O.}\ \bibnamefont
			{Powell}},\ }\bibfield  {title} {\bibinfo {title} {Growth {Rate} and
			{Generation} {Time} of {Bacteria}, with {Special} {Reference} to {Continuous}
			{Culture}},\ }\href {https://doi.org/10.1099/00221287-15-3-492} {\bibfield
		{journal} {\bibinfo  {journal} {J. Gen. Microbiol.}\ }\textbf {\bibinfo
			{volume} {15}},\ \bibinfo {pages} {492} (\bibinfo {year} {1956})}\BibitemShut
	{NoStop}%
	\bibitem [{\citenamefont {Hashimoto}\ \emph {et~al.}(2016)\citenamefont
		{Hashimoto}, \citenamefont {Nozoe}, \citenamefont {Nakaoka}, \citenamefont
		{Okura}, \citenamefont {Akiyoshi}, \citenamefont {Kaneko}, \citenamefont
		{Kussell},\ and\ \citenamefont {Wakamoto}}]{hashimoto_noise-driven_2016}%
	\BibitemOpen
	\bibfield  {author} {\bibinfo {author} {\bibfnamefont {M.}~\bibnamefont
			{Hashimoto}}, \bibinfo {author} {\bibfnamefont {T.}~\bibnamefont {Nozoe}},
		\bibinfo {author} {\bibfnamefont {H.}~\bibnamefont {Nakaoka}}, \bibinfo
		{author} {\bibfnamefont {R.}~\bibnamefont {Okura}}, \bibinfo {author}
		{\bibfnamefont {S.}~\bibnamefont {Akiyoshi}}, \bibinfo {author}
		{\bibfnamefont {K.}~\bibnamefont {Kaneko}}, \bibinfo {author} {\bibfnamefont
			{E.}~\bibnamefont {Kussell}},\ and\ \bibinfo {author} {\bibfnamefont
			{Y.}~\bibnamefont {Wakamoto}},\ }\bibfield  {title} {\bibinfo {title}
		{Noise-driven growth rate gain in clonal cellular populations},\ }\href
	{https://doi.org/10.1073/pnas.1519412113} {\bibfield  {journal} {\bibinfo
			{journal} {Proc. Natl. Acad. Sci. U.S.A.}\ }\textbf {\bibinfo {volume}
			{113}},\ \bibinfo {pages} {3251} (\bibinfo {year} {2016})}\BibitemShut
	{NoStop}%
\end{thebibliography}



%

\end{document}